\documentclass[12pt]{article}

\usepackage{bigstrut}
\usepackage{comment}
\usepackage[a4paper, total={6in, 8in}]{geometry}
\usepackage{authblk}
\title{Connections between propulsive efficiency and wake structure via modal decomposition}
\author{Morgan R. Jones$^*$}
\author{Eva Kanso}
\author{Mitul Luhar}
\date{}
\affil{Department of Aerospace and Mechanical Engineering, University of Southern California, Los Angeles, CA 90010}
\usepackage{amsmath}
\usepackage{verbatim}
\usepackage{hyperref}
\usepackage{ragged2e}

\usepackage[sc]{mathpazo}
\usepackage[T1]{fontenc}

\hypersetup{
    colorlinks=true,
    linkcolor=blue,
    filecolor=magenta,      
    urlcolor=blue,
    citecolor=blue,
}
\usepackage{xcolor}
\usepackage{caption}

\usepackage{natbib}
\setcitestyle{authoryear}

\usepackage{graphicx}
\usepackage{float}
\graphicspath{ {./images/} }

\begin{document}\sloppy
\maketitle

\begin{abstract}
We present experiments on oscillating hydrofoils undergoing combined heaving and pitching motions, paying particular attention to connections between propulsive efficiency and coherent wake features extracted using modal analysis. Time-averaged forces and particle image velocimetry (PIV) measurements of the flow field downstream of the foil are presented for a Reynolds number of $Re=11\times10^3$ and Strouhal numbers in the range $St=0.16-0.35$. These conditions produce 2S and 2P wake patterns, as well as a near-momentumless wake structure. A triple decomposition using the optimized dynamic mode decomposition (opt-DMD) method is employed to identify dominant modal components (or coherent structures) in the wake. These structures can be connected to wake instabilities predicted using spatial stability analyses. Examining the modal components of the wake provides insightful explanations into the transition from drag to thrust production, and conditions that lead to peak propulsive efficiency. In particular, we find modes that correspond to the primary vortex development in the wakes. Other modal components capture elements of bluff body shedding at Strouhal numbers below the optimum for peak propulsive efficiency and characteristics of separation for Strouhal numbers higher than the optimum.
\end{abstract}
\begin{keywords} 
\smallskip low-dimensional models, swimming/flying, jets/wakes
\end{keywords}

\section{Introduction}
The performance characteristics of swimming and flying animals have long motivated the design of autonomous swimmers with similar kinematics for propulsion and navigation. The energetics of fish propulsion and, in particular, their ability to harvest energy in schools or unsteady flows environments \citep{beal_passive_2006} have also spurred the research and development of novel energy harvesters \citep{mckinney_wingmill_1981,jones_numerical_1997,bryant_wake_2012}. Broadly, these animals rely on oscillating or undulating foils for propulsion. 

Characterizing the relationship between foil performance and the downstream wake structure has been a long-standing research area. Pioneering work in this area by \cite{triantafyllou_optimal_1993} analyzed the swimming performance of several different fish and showed that fish operate within a narrow Strouhal number range, $0.2 < St < 0.35$, that leads to high propulsive efficiency. The Strouhal number is a kinematic parameter defined as $St=f A/U_{\infty}$, where $A$ is a characteristic length describing the width of the wake, $f$ is the frequency of oscillation, and $U_{\infty}$ is the swimming speed. Laboratory experiments also showed high propulsive efficiencies from oscillating foils operating in this Strouhal number range. Complementary stability analyses showed that oscillating foils produce a jet-like wake that is convectively unstable when excited at frequencies corresponding to Strouhal numbers that yield peak propulsive efficiencies. The spatial stability analysis performed by \cite{triantafyllou_optimal_1993} used the jet profile of a reverse von K\'{a}rm\'{a}n vortex street, where two single (S) vortex cores were shed per half cycle of oscillation, i.e., a 2S wake pattern. Similar relationships between optimal Strouhal number ranges and instabilities of the jet wake were found by \cite{lewin_modelling_2003}. 

The relationship between wake instability and peak propulsive efficiency was studied further by \cite{moored_hydrodynamic_2012}, who used Particle Image Velocity (PIV) to visualize wake structures produced by a batoid-inspired oscillating fin. These experiments showed multiple peak efficiencies, with some corresponding to a 2P wake pattern, where two pairs (P) of vortex cores are shed per half cycle, and others corresponding to a 2S wake. \cite{moored_hydrodynamic_2012} also pursued local stability analyses (i.e., relying on a parallel flow assumption) to show that the time-averaged velocity profile exhibits instabilities when excited at frequencies corresponding to peak propulsive efficiency. This phenomenon was termed 'wake resonance', suggesting that when the oscillating foil is tuned to optimally excite the wake, it produces the highest propulsive efficiencies.  
\cite{moored_hydrodynamic_2012} also examined the vorticity perturbations generated by the most unstable modes in an effort to better understand these instabilities. Follow-on work by \cite{moored_linear_2014} showed that the 'resonant frequencies' correspond to optimal momentum entrainment, both into and out of the jet region. These findings suggest that there is no single wake structure that corresponds to peak propulsive efficiency \citep{smits_undulatory_2019}. Further examples of 2P vortex patterns have also been found in wakes produced by eels and dolphins \citep{tytell_hydrodynamics_2004, smits_undulatory_2019}.

Recent studies have raised questions regarding the validity of wake resonance theory \citep{arbie_stability_2016}, noting that although correlations exist between unstable frequencies and peak propulsive efficiency from experiments, it is more difficult to establish a causal link between the two. \cite{arbie_stability_2016} considered the stability characteristics of momentumless wakes and noted that these wakes may be stable (even if the thrust producing jet-like component is unstable). 
It has also been suggested that, while wake structure provides insight into propulsive efficiency, it cannot provide a complete explanation \citep{zhang_footprints_2017,taylor_simple_2018}.
Other studies \citep{eloy_optimal_2012} suggest that the development of the wake structure is a result of, and not a cause of, high propulsive efficiency. The reverse von K\'{a}rm\'{a}n wake from an oscillating foil can be shed in more patterns than just the 2S and 2P wake \citep[see e.g.,][]{lentink_vortex-wake_2008, schnipper_vortex_2009,andersen_wake_2017}, and each wake structure has an indirect relationship with propulsive efficiency. For example, \cite{mackowski_williamson_2015} found from experiments that wake patterns for a pitching foil, where self-interactions of the vortices occur, do not reflect a net force. Thus, observing the wake structure alone is likely not enough to completely determine the propulsive efficiency \citep{floryan_swimmers_2020}. 

As an alternative approach, we make use of dynamic mode decomposition (DMD) to identify coherent features (or patterns) in the wakes produced by oscillating foils and link these features to propulsive performance. DMD, as first introduced by \cite{schmid_dynamic_2010}, is a technique used to approximate the dynamics of a nonlinear system via the identification of a linear operator that evolves the system to the next state. DMD can also be thought of as a data-driven stability analysis because each DMD mode is associated with a specific frequency and growth or decay rate, which provides interpretable physical insight into the spatial structures and their dynamic evolution. In contrast to the stability analysis approach, DMD does not require the mean flow to be locally or globally parallel. In the fields of marine propulsion and energy harvesting, DMD and similar modal decomposition techniques have been used on propeller and turbine wakes to characterize wake instabilities, loading conditions, and efficiency \citep{magionesi_modal_2018, sarmast_mutual_2014, strom_near-wake_2022,araya_transition_2017}.

In this study, we use the triple decomposition to identify coherent flow features that contribute to drag and thrust production from PIV measurements for the wake past an oscillating foil. We examine both the vorticity and the Reynolds stresses associated with optimized DMD modes across Strouhal numbers ranging from $St = 0.16$ to $St = 0.35$, and compare the results to time-averaged forces and propulsive efficiencies. We find that the use of DMD enables us to link wake structure with wake stability and propulsive efficiency for oscillating foils. 

Our study builds upon the foil configuration described by \cite{floryan_swimmers_2020}. We focus on rigid foils and present results primarily in the context of swimming performance. While the use of rigid foils represents a simplification of the fluid-structure interactions pertinent to flapping foil propulsion in nature, our results may provide additional insight into coherent flow features that contribute to drag and thrust.  Moreover, the observations presented herein are also relevant to the design of autonomous swimming vehicles \citep{buren_floryan_smits_2020} and energy harvesting systems \citep{mckinney_wingmill_1981} that use rigid oscillating foils.
\section{Heave and pitch foil parameters}
\begin{figure}[t!]
\centering
\includegraphics[width=0.5\textwidth,keepaspectratio]{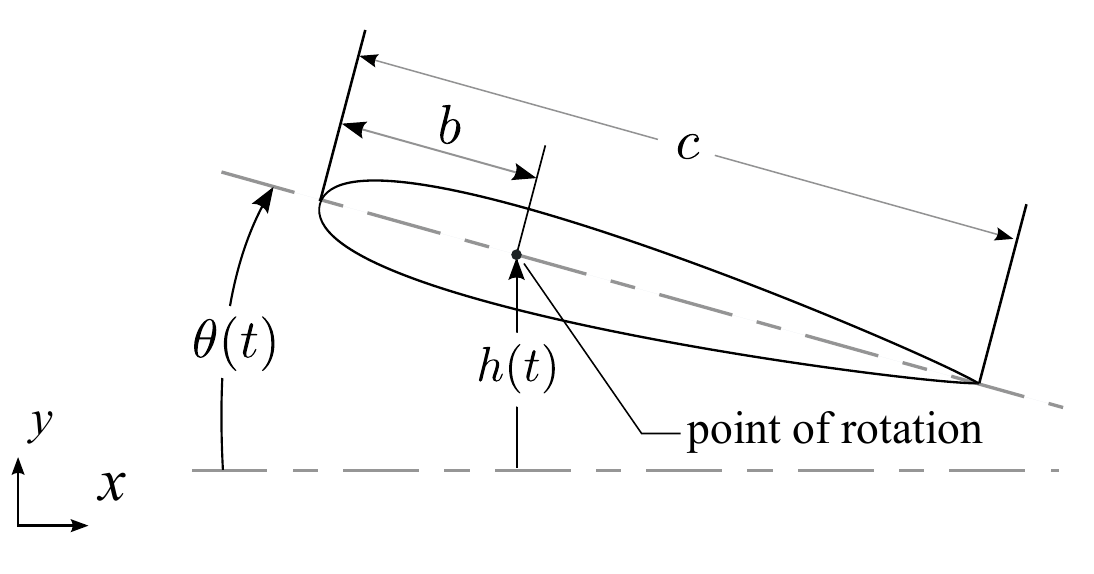}
\caption{Geometry and parameters for a pitching and heaving hydrofoil.}
\label{fig:geometry}
\end{figure}

We consider an oscillating NACA-0012 hydrofoil with chord $c$ and span $s$ as illustrated in figure \ref{fig:geometry}. The distance $b=0.25c$ denotes the point of rotation from the leading edge. The imposed pitching and heaving kinematics of the foil at the point of rotation are described by $\theta(t) =\theta_0\sin(2\pi f t+\phi_p)$ and $ h(t) =h_0\sin(2\pi f t)$ respectively, where $ \theta_0 $ and $ h_0 $ are the respective pitching and heaving amplitudes, $ f $ is the frequency of oscillation and $ \phi_p $ is the phase difference between heaving and pitching oscillations. For the experiments described below, the phase difference was set to $\pi/2$ ($90^\circ$). 

The oscillation frequency can be expressed in dimensionless terms as the Strouhal number:
\begin{equation}
   St =\frac{f A_{TE}}{U_\infty}.
\end{equation}
Note that the Strouhal number can also be interpreted as the wake width ($\sim A_{TE}$) normalized by the wavelength ($\sim U_{\infty}/f$).  Another important parameter relative to swimming performance is the angle of attack $\alpha$, which is different from the pitch angle due to the influence of the heave velocity, $\dot{h}(t)$, and can be expressed as:
 \begin{equation}
   \alpha (t) =\arctan{\left(\frac{1}{U_{\infty}}\dot{h}(t)\right)}-\theta(t).
\end{equation}
In the case where the phase angle $\phi_p$ between pitch and heave is $90^\circ$, the maximum angle of attack $\alpha_0$ can be estimated as:
\begin{equation}
    \alpha_0=\arctan\left(\frac{\omega h_0}{U_{\infty}}\right)-\theta_0,
\end{equation}
where $\omega = 2\pi f $ is the radian frequency. 
The time averaged power and thrust coefficients are described using the following equations:
\begin{equation}\label{eq:CP_CT}
   C_P =\frac{\bar{\wp}}{\frac{1}{2}\rho s c U_{\infty}^3}, \ C_T =\frac{\bar{\tau}}{\frac{1}{2}\rho s c U_{\infty}^2},
\end{equation}
where $\rho$ is the density of the fluid, $\bar{\tau}$ is the time-averaged thrust, and $\bar{\wp}=\frac{1}{T}\left(\int_{0}^{T} \dot{h}(t) F_y(t) dt + \int_{0}^{T} \dot{\theta}(t)T_z(t) dt\right)$ is the time-averaged power input to the fluid. $F_y$ and $T_z$ are the force and torque associated with the heaving and pitching motions. Propulsive efficiency can then be calculated as:
\begin{equation}\label{eq:eta}
   \eta =\frac{C_T}{C_P} = \frac{\bar{\tau} U_\infty}{\bar{\wp}}.
\end{equation}
The propulsive efficiency and thrust coefficient are both used to characterize swimming performance. Table~\ref{table:1} shows kinematic parameters and flow conditions examined in the present study. Note that we match kinematic parameters with those from \cite{QuinnOpt}, who presented a large dataset of propulsive efficiencies for a NACA-0012 foil. 

\begin{center}
\centering
\begin{table}
\resizebox{\textwidth}{!}{\begin{tabular}{ c c c c c c c c} 
 \hline
 \\
 Study & Configuration & Type & $Re_c \ (\times 10^3)$ & $St$  & $\phi_p$ & $\theta_0$ & $h^*$\\
 
 Present Study & Pitch and Heave & E & 11 & 0.16$-$0.59 & $90^\circ$ & $10^\circ$ & 0.19 \\ 
 \cite{QuinnOpt} & Pitch and Heave & E & 5$-$70 & 0.14$-$1.4 & $90^\circ$ & $10^\circ$ & 0.19\\
\end{tabular}}
\caption{Parameters used in the present and previous oscillating-foil studies. Chord-based Reynolds number $Re_c=\rho U_\infty c/\mu$, Strouhal Number, $St$, phase between heaving and pitching $\phi_p$, pitch amplitude $\theta_0$, and dimensionless heave ratio $h^*=h_0/c$.}
\bigstrut \hrule
\label{table:1}
\end{table}
\end{center}

\section{Experimental Methods}
\begin{figure}[t!]
\centering
\includegraphics[width=0.75\textwidth]{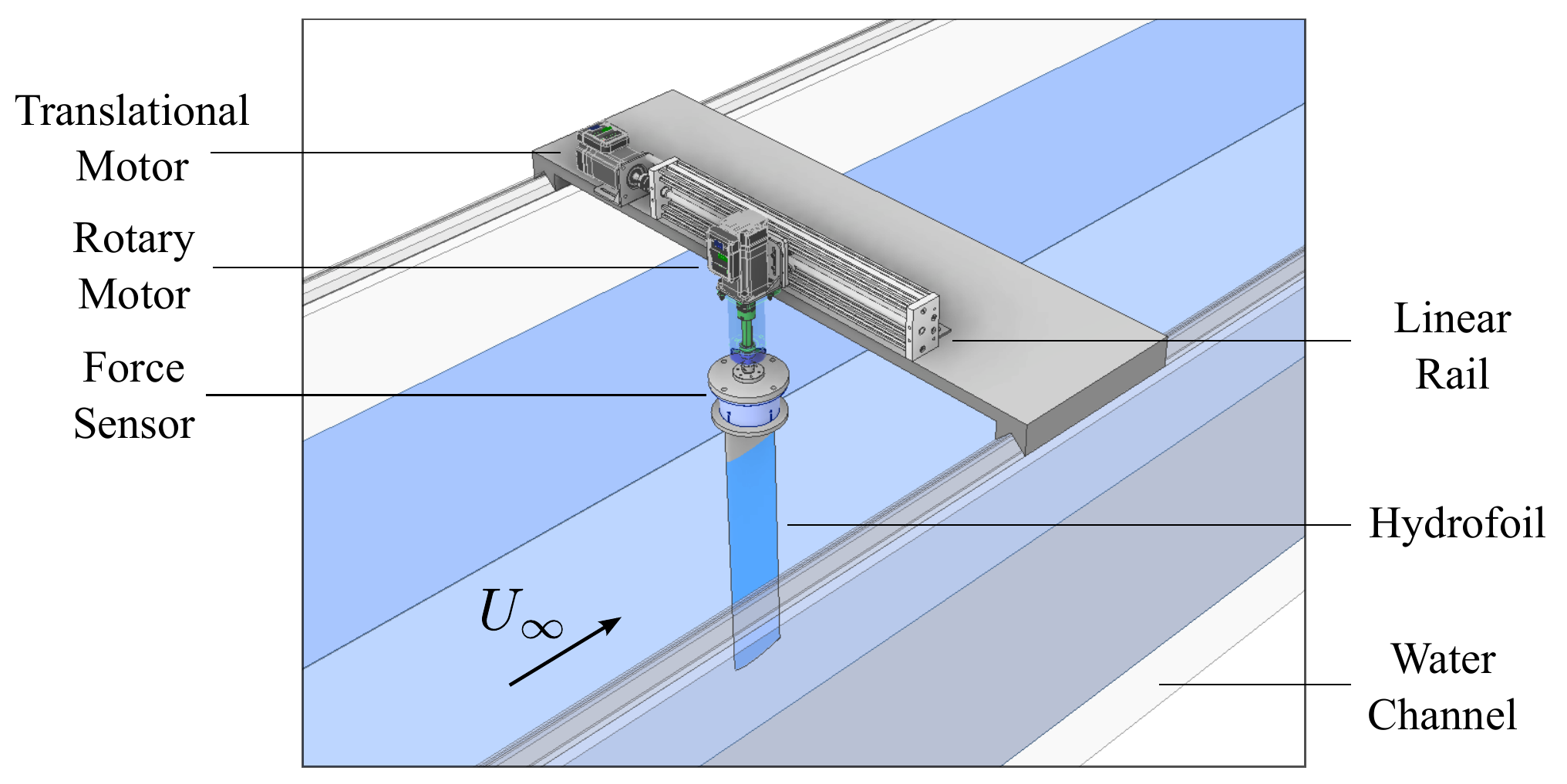}
\caption{Experimental motion-control system.}
\label{fig:setup}
\end{figure}

Experiments were carried out in a large-scale free-surface water channel with a glass-walled test section of dimensions 7.6 m $\times$ 0.6 m $\times$ 0.9 m. A NACA-0012 hydrofoil with a chord of $c=10$ cm and a span of $s=32$ cm was used. The foil was 3D printed from polylactic acid (PLA) filament, sanded, and reinforced with two internal aluminum rods throughout the span. The channel was run at a freestream speed of $U_\infty = 0.1$ m/s with a turbulence intensity of 1$\%$. Foil motions were produced using a closed-loop control system driven by two NEMA 23 integrated stepper motors, as shown in figure \ref{fig:setup}. One of the motors was connected to a linear rail to generate the heaving motion. The other motor was mounted to the beam to create the pitching motion. Translational and angular positions of the foil were calculated from the precision micro-steps (3200 steps per revolution) of the stepper motors. Measurements of hydrodynamic forces were made concurrently with an ATI Gamma (SI-32-2.5) six-degree-of-freedom force transducer with a minimum force and torque resolution of $1/80$ N and $1/2000$ Nm respectively. Data were acquired at a rate of 5 kHz and filtered using a zero-phase second-order Butterworth filter. Instantaneous forces were measured over oscillation frequencies $f=0.4$ Hz$ - 1.3$ Hz. For the kinematic parameters used in the present study (Table~\ref{table:1}), this yielded a Strouhal number range of $St=0.16 - 0.59$. Time-averaged thrust coefficients ($C_T$), power coefficients ($C_P$), propulsive efficiencies ($\eta$) were computed by combining the force measurements with the foil kinematic data. The standard Klein-McClintock procedure was used to estimate uncertainties for these parameters based on known instrument resolution and measurement uncertainty for the time-averaged thrust and power input \citep{taylor1997introduction}. Force measurements were collected and averaged for at least 13 periods for each run. Freestream velocity measurements were recorded using a Laser Doppler Velocimetry system (MSE miniLDV). Relative uncertainties in the velocity measurements were $0.11\%$.

2D-2C PIV measurements were carried out in the near-wake region of the oscillating foil to provide further insight into the transition from drag-producing and free-swimming conditions to thrust-producing conditions. The PIV system comprised a 5 W 532 nm solid-state laser and a Phantom VEO 410-L high speed camera with $1280 \times 800$ pixel resolution. Images of the flow field were captured in a field-of-view of $27$ cm $\times$ $17$ cm at a rate of $65$  Hz for approximately $56$ s. Standard analysis routines in DaVis (LaVision) were used to generate velocity measurements. Two $64$ pixel $\times$ $64$ pixel and four $24$ pixel $\times$ $24$ pixel convolution windows were used with a $50\%$ overlap to determine the velocity vectors. This yielded a set of $77 \times 50$ velocity vectors and 3600 snapshots for each of the wakes in this study. Erroneous columns of velocity values from the far ends of the field-of-view were removed.  

\begin{figure}[t!]
\centering
\includegraphics[width=0.6\textwidth]{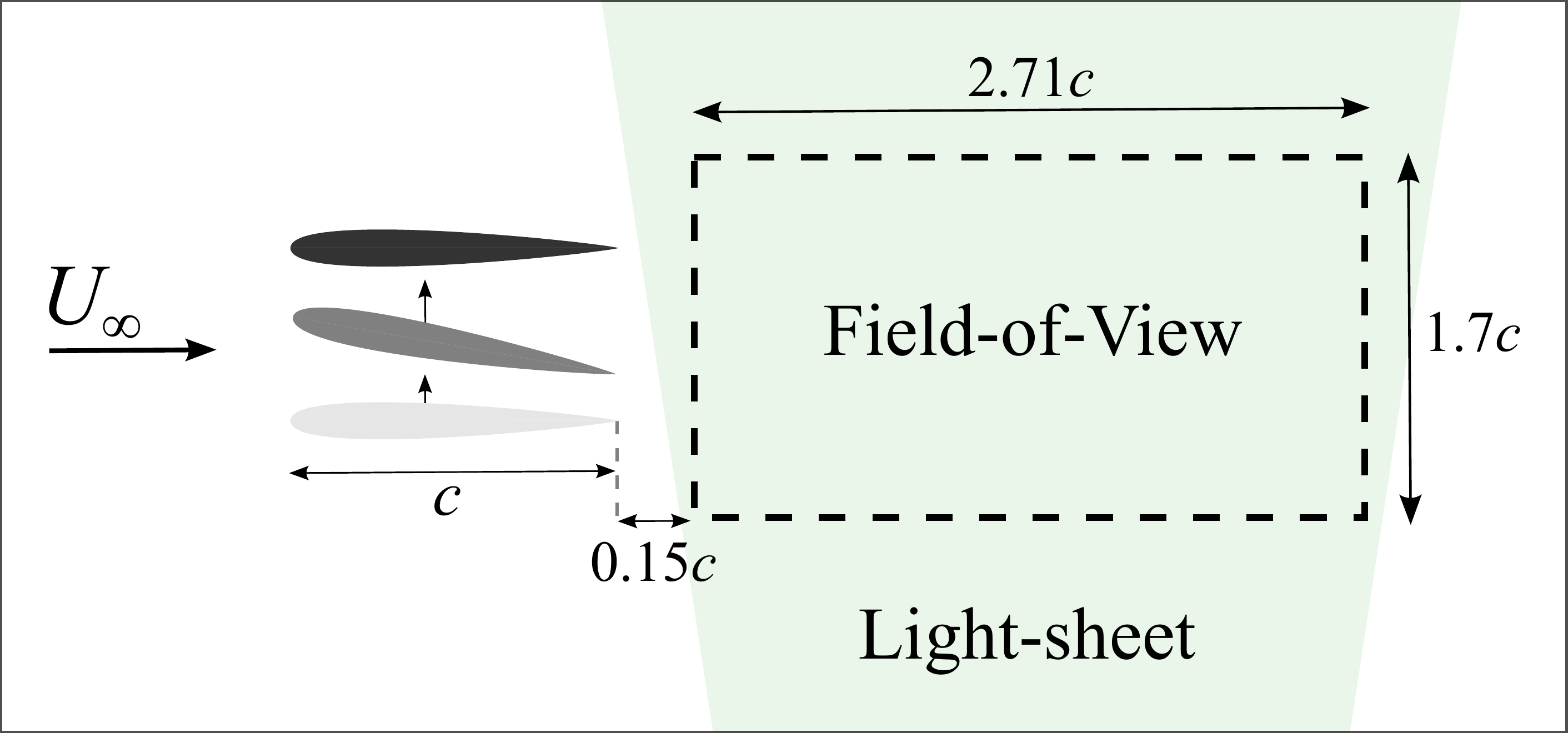}
\caption{Hydrofoil and PIV experimental setup with field-of-view dimensions.}
\label{fig:fieldofView}
\end{figure}

PIV flow fields were both phase-averaged and time-averaged. For the lowest oscillation frequency ($f=0.4$ Hz), the available PIV data spanned more than 21 oscillation periods. Dimensions of the field-of-view, relative to the hydrofoil are illustrated in figure~\ref{fig:fieldofView}. Time- and phase-averaged measurements of streamwise velocity and vorticity were computed for the Strouhal numbers $St= 0.16$, $0.23$, $0.29$, and $0.35$. The respective maximum angles of attack were $\alpha_0=$ $15.2^\circ$, $22.3^\circ$, $28.2^\circ$, and $33.1^\circ$.

\subsection{Periodic Structures via the Triple Decomposition}
For periodic flows under natural or forced conditions, large-scale coherent motions are often present in addition to turbulence. In such situations, the Reynolds decomposition, which assumes that turbulence is the only source of fluctuations in the flow, may not be accurate and can lead to overestimation of the stochastic part of the flow. Alternatively, the full flow field $\mathbf{u}$ can be expressed as a triple decomposition as introduced by \cite{hussain_mechanics_1970}:
\begin{equation} \label{eq:triple}
   \mathbf{u}(\mathbf{x},t)=\mathbf{\bar{u}(x)}+\mathbf{\tilde{u}}(\mathbf{x},\phi(t))+\mathbf{u'}(\mathbf{x},t), 
\end{equation}

where $\mathbf{\bar{u}}$ is the mean (time-averaged) flow field, $\mathbf{\Tilde{u}}$ is the periodic flow field, characterized by the phase parameter $\phi(t)$, and $\mathbf{u'}$ represents the turbulent fluctuations that are incoherent. In our case, where the flow is driven by a known forcing frequency, the periodic component can be directly obtained through phase-averaging as:
\begin{equation} \label{eq:periodic}
    \mathbf{\tilde{u}}(\mathbf{x},\phi_0)+\mathbf{\bar{u}}(\mathbf{x})=\frac{1}{N}\sum_{n=1}^N \mathbf{u}(\mathbf{x},\phi_0),
\end{equation}
where the specific phase position $\phi_0=\phi(t_0+n\tau)$ is defined for an initial time $t_0$ and a period $\tau$. However, for natural flows, or those driven by a series of forcing frequencies, equation \ref{eq:periodic} may be difficult to compute. Instead, phase averaging can be attained through modal decomposition methods. In these cases, the unsteady coherent component $\mathbf{\Tilde{u}}$ may be represented by a linear combination of modes obtained via techniques such as Proper Orthogonal Decomposition (POD) or Dynamic Mode Decomposition (DMD), and the triple decomposition can be modified as described by \cite{baj_triple_2015}:
\begin{equation} \label{eq:tripleDMD}
   \mathbf{u}(\mathbf{x},t)=\mathbf{\bar{u}(x)}+\sum_{n=1}^r\mathbf{\tilde{u}}_n(\mathbf{x},\phi_n(t))+\mathbf{u'}(\mathbf{x},t).
\end{equation}

Here, $r$ is the number of modes retained from the modal decomposition. 
In this study, we use Dynamic Mode Decomposition to obtain the periodic component of the flow field in equation \ref{eq:tripleDMD} and compare this modal representation to the wake characteristics obtained via direct phase-averaging, as shown in equation \ref{eq:periodic}.

\subsection{Dynamic Mode Decomposition}
DMD was originally presented to the fluid mechanics community by \cite{schmid_dynamic_2010} as a method to decompose unsteady or turbulent flow fields into coherent structures. DMD can be interpreted as an eigendecomposition of a least squares best-fit linear operator $\mathbf{A}$ that advances the past snapshots of data (i.e., 2D-2C velocity fields obtained from PIV) forward in time towards future snapshots.
For a set of $m$ snapshots ($\mathbf{x}_1, \mathbf{x}_2, \cdots , \mathbf{x}_m$), the dataset is arranged into two separate matrices of the following forms:
\begin{equation}
\mathbf{X}=\begin{bmatrix}
\vert & \vert & & \vert\\
\mathbf{x}_1 & \mathbf{x}_2 & ... & \mathbf{x}_{m-1}\\
\vert & \vert & & \vert
\end{bmatrix},
\end{equation}
and
\begin{equation}
\mathbf{X}'=\begin{bmatrix}
\vert & \vert & & \vert\\
\mathbf{x}_2 & \mathbf{x}_2 & ... & \mathbf{x}_{m}\\
\vert & \vert & & \vert
\end{bmatrix},
\end{equation}
with
\begin{equation}
   \mathbf{X'= A X}. 
\end{equation}

Generally, $\mathbf{A}$ can be approximated by taking the Moore-Penrose pseudo inverse of $\mathbf{X'} \ (\mathbf{A=X' X^{-1})}$ via the singular value decomposition (SVD): $\mathbf{X=U \Sigma V^*}$. However, for high-dimensional datasets such as measurements from PIV snapshots, the computation of $\mathbf{A}$ is expensive and can be inaccurate due to noisy entries or outliers in the data. To limit computational expense and the impact of noise, we consider the projection of $\mathbf{A}$ onto a reduced-rank representation of the data obtained via the SVD, $\mathbf{X \approx U_r \Sigma_r V_r^*}$, where the subscript denotes a truncation to rank-$r$:
\begin{equation}
    \mathbf{\Tilde{A}=U^*_r A U_r = U^*_r X'V_r \Sigma^{-1}_r}.
\end{equation}
Note that this is analogous to projecting $\mathbf{A}$ onto the leading POD modes. Subsequently, the eigenvalues of matrix $\mathbf{A}$ can be estimated by taking the eigendecomposition of $\mathbf{\Tilde{A}}$:
\begin{equation}
    \mathbf{\Tilde{A} W=W \Lambda},
\end{equation}
where the matrix $\mathbf{W}$ contains the eigenvectors and $\mathbf{\Lambda}$ contains the corresponding eigenvalues ($\lambda_1, \lambda_2, \cdots$). Dynamic modes for the flow field can be computed using the exact DMD algorithm from \cite{tu_dynamic_2014} as:
\begin{equation}
    \mathbf{\Phi=X' V_r \Sigma_r^{-1} W_r}.
\end{equation}
Here the matrix $\mathbf{\Phi}$ contains individual dynamic modes $(\phi_1, \phi_2, \cdots)$.
Using this low-dimensional approximation to the dynamics represented by $\mathbf{A}$, we can reconstruct the data using a linear approximation of the dynamic modes for all future times:
\begin{equation}
    \mathbf{x}(t)\approx \sum_{k=1}^r \phi_k\exp(\omega_k t)b_k = \boldsymbol{\Phi} \exp(\boldsymbol{\omega} t)\mathbf{b}.
\end{equation} 
The frequency $\omega_k$ is defined based on the associated eigenvalue $\lambda_k$ as $\omega_k=\ln(\lambda_k)/\Delta t$, where $\Delta t$ is the time interval between individual snapshots. The vector $\mathbf{b}$ contains entries of the initial amplitudes $b_k$ defined as $\mathbf{b}=\mathbf{\Phi}^{-1}\mathbf{x_1}$. Recall that the dynamic modes $\phi_k$ are the columns of the matrix $\mathbf{\Phi}$. 
The modes in this temporal DMD formulation represent the absolute stability of the flow field, with the eigenvalues providing insight into oscillation frequencies and growth or decay rates. Given a complex eigenvalue $\lambda_k = \lambda_{k,r} + i \lambda_{k,i} = |\lambda_k| \exp(i \angle \lambda_k)$, the frequency can be estimated as $\omega_k = \omega_{k,r} + i \omega_{k,i} = \frac{\ln|\lambda_k|}{\Delta t} + i \frac{\angle \lambda_k}{\Delta t}$. Thus, a dynamic mode $\phi_k$ grows over time if the corresponding eigenvalue has amplitude $|\lambda_k|>1$ and decays if $|\lambda_k|<1$.

Spatial growth rates can also be approximated by the DMD algorithm by reorganizing the series of snapshots as increasing in space. In this case, modes in spatial DMD represent convective stability, where the eigenvalues would represent spatial frequencies or wavenumbers. In practice, spatial DMD is more prone to noise due to the sparsity of spatial data from PIV measurements, since the number of time-snapshots available is typically much higher than the spatial locations \citep{schmid_dynamic_2010}. For instance, the PIV data obtained in this study span 3600 snapshots in time but only 77 streamwise locations for each run.  As a result, we focus on using the temporal DMD approach to take advantage of our time-resolved snapshot data.

\subsection{Optimized DMD}
For the periodic flows that are studied in this paper, it is expected that the dynamic modes should neither grow or decay in time. That is, the eigenvalues of DMD are fixed directly onto the unit circle $|\lambda_k|=1$. However, the presence of even weak noise in periodic flows is known to yield damped eigenvalues, $|\lambda_k|<1$ that result in decay of the corresponding modes ($\phi_k$) over time \citep{bagheri_effects_2014}. Thus, several recent methods have been proposed for debiasing the DMD algorithm in the presence of noisy data, such as measurements from PIV \citep{dawson_characterizing_2016,hemati_-biasing_2017, askham_variable_2018}. For this purpose we use the optimized-DMD (opt-DMD) algorithm presented by \cite{askham_variable_2018}. This is a variant of DMD that uses a variable projection method to approximate the linear operator $\mathbf{A}$ with reduced noise. Specifically, opt-DMD solves the non-linear least-squares problem:
\begin{equation} \label{eq:optdmd}
    \underset{\Lambda, \mathbf{b}}{\min}||\mathbf{X}^T - \mathbf{\Phi}(\Lambda)\mathbf{b}||_F,
\end{equation}
where the eigenvalues in $\Lambda$ act as iterative parameters and determine the values of the eigenfunctions, $\mathbf{\Phi}$, and amplitude coefficients, $\mathbf{b}$. The algorithm also has the advantage of projecting the modes onto the full dataset rather than a subset from an $r$-rank truncation. Additional constraints can be applied to the modes, such as restricting eigenvalues to stay on the unit circle. For simplicity, we do not use any constraints for the flow fields in this study, and find that opt-DMD naturally converges towards the expected eigenvalues in a periodic flow. 
The DMD modes of velocity are computed using both opt-DMD and exact DMD algorithm for comparison. We compute the vorticity directly from the opt-DMD modes of velocity to understand the differences between the modal representation and the full flow field. 

\section{Results}
\subsection{Propulsive Efficiency}
\begin{figure}[t!]
\centering
\includegraphics[width=1\textwidth]{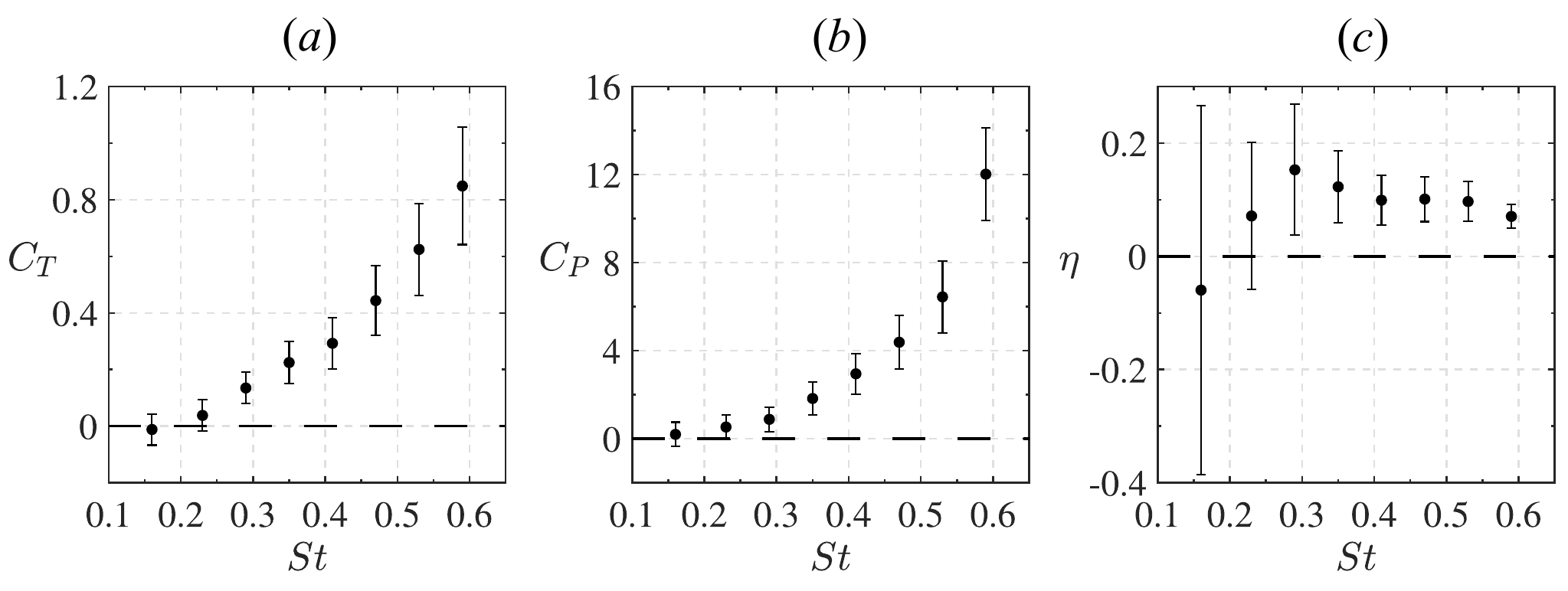}
\caption{Comparison of time-averaged thrust coefficients and propulsive
efficiencies as a function of Strouhal number.}
\label{fig:Forces}
\end{figure}

Measurements of the thrust coefficient and propulsive efficiency are presented in figure \ref{fig:Forces}. The thrust and power coefficients increase monotonically with increasing Strouhal number. However, $C_P$ increases faster than $C_T$ at higher $St$ values, leading to non-monotonic behavior in propulsive efficiency and reduced efficiency for $St > 0.35$. Peak propulsive efficiencies can be seen within the range of $0.23<St<0.35$. With the provided uncertainty estimates, it can be argued that either of the three Strouhal numbers $St=0.23$, $0.29$, and $0.35$ result in the highest propulsive efficiency. However, simplified scaling laws suggest that for angles of attack $\alpha$ where the flow remains attached to the foil, peak propulsive efficiency can be estimated as 
\begin{equation} \label{taylorSt}
    St_{max}\approx\sqrt{3}St_0,
\end{equation} 
where $St_0$ is the Strouhal number that results in zero net thrust \citep{taylor_simple_2018}. In our case, the thrust coefficient is near zero for $St=0.16$. Using this value for $St_0$ gives a maximum efficiency of $St_{max}\approx0.28$. This approximation is consistent with our case of $St=0.29$ for which the measured $\eta$ is highest. Note that the uncertainty in measured efficiency at $St=0.16$ is large, which is due to the fact that the thrust and power coefficients are close to $0$ for this condition. Nonetheless, this case is similar to a \textit{self-propelled} swimming mode, where the thrust from the foil is approximately equal to its drag. Negative efficiencies are expected for lower Strouhal numbers, $St<0.16$, for which the net thrust becomes negative due to the effects of viscous drag. As the foil oscillates at lower frequencies, the drag contribution on the propulsor stays approximately constant while the thrust generated decreases \citep{Floryan}. The present measurements are broadly consistent with trends observed from previous studies \citep{QuinnOpt,triantafyllou_optimal_1993}.

\subsection{Mean and Phase-averaged Flow Fields} 
\label{meanphase}

\begin{figure}[t!]
\begin{center}
\includegraphics[width=0.85\textwidth]{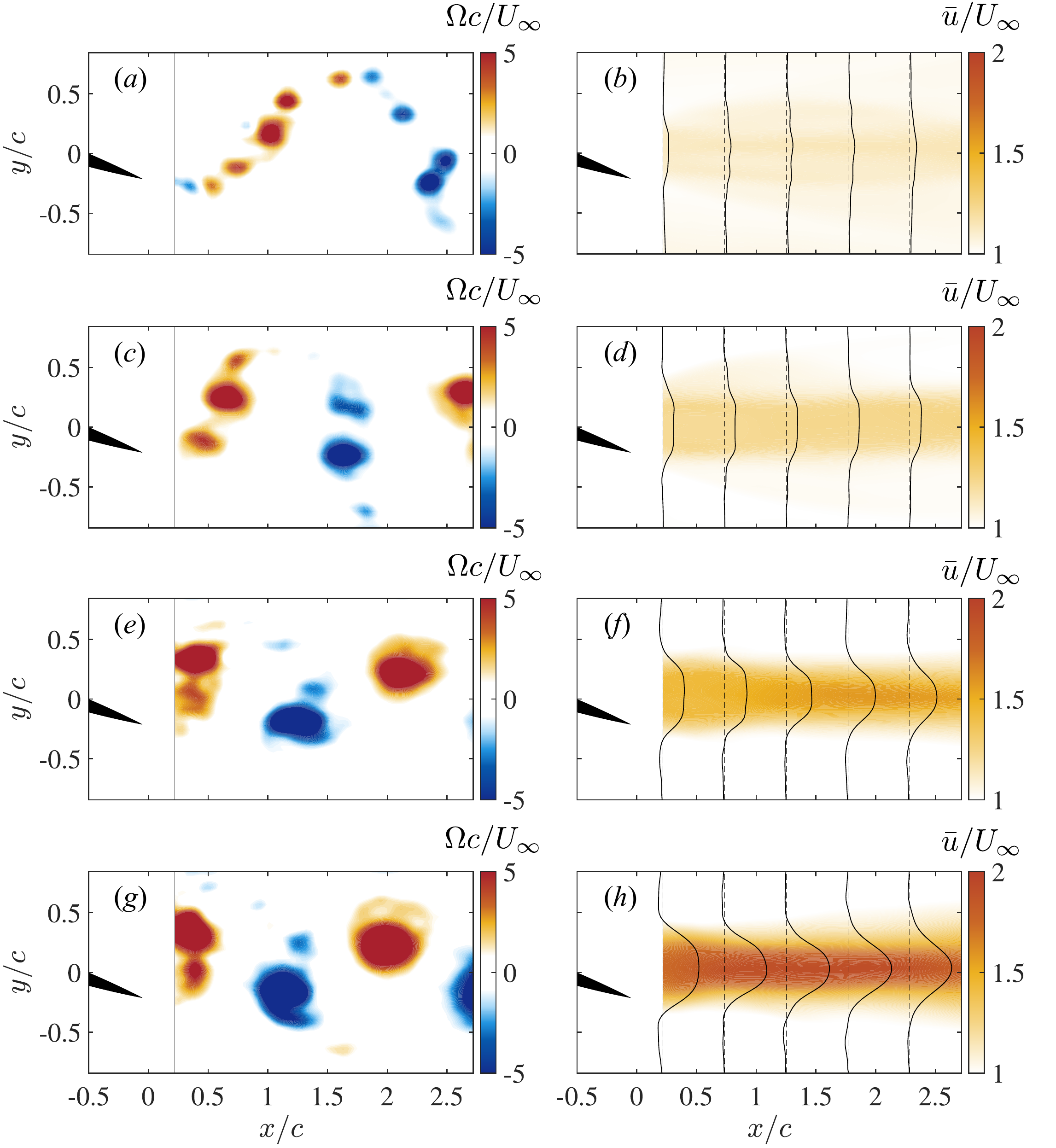}
\caption{Phase averaged vorticity and time averaged stream-wise velocity fields for ($a-b$) $St=0.16$, ($c-d$) $St=0.23$, ($e-f$) $St=0.29$, and ($g-h$) $St=0.35$. See also supplementary movie 1.}
\label{fig:VortPIV}
\end{center}
\end{figure}

PIV measurements were used to study the downstream wake of the foil for a subset of the Strouhal numbers for which the force measurements were obtained. Phase-averaged vorticity ($\Omega$) and time-averaged streamwise velocity ($\bar{u}$) fields are shown in in figure \ref{fig:VortPIV}, where the foil is positioned to move in the positive $y$ direction. An increase in magnitude of the \textit{jet} wake (i.e., $\bar{u}/U_\infty > 1$) can be seen with increasing $St$. A near-momentumless wake is observed for $St=0.16$, where the wake profiles are close to zero, again indicating a resemblance to self-propelled swimming (figure \ref{fig:VortPIV}$b$). Under this condition, approximately five positive and six negative vortices are shed per cycle of oscillation, signifying a small wake asymmetry. The vortices deflect laterally away from the heave and pitch centerline, resulting in an increase of the wake width from $x/c \approx 0.5 - 1.5$ that can be seen in figure \ref{fig:VortPIV}$b$. 

Increasing the Strouhal number to $St=0.23$ produces a 2P wake (figure \ref{fig:VortPIV}$c$), with mean profiles that are approximately trapezoidal-shaped, with two small but distinct peaks. This wake characteristic matches well with the 2P structure observed by \cite{dewey_relationship_2012}, who concluded that the presence of two vortex pairs result in two separate peaks in the mean profile. The Strouhal number $St=0.29$ (figure \ref{fig:VortPIV}$e$), initially produces a 2P wake with mean profiles that are trapezoidal ($0.22\le x/c \le 1.5$). These vortex pairs coalesce further downstream to a 2S wake, creating a mean profile resembling that produced by a single jet ($1.5 < x/c \le 2.7$). As the Strouhal number increases further to $St = 0.35$, a classical 2S wake is observed, suggesting a transition from 2P to 2S wake structure over the range $0.23<St<0.35$. These observations are consistent with the results of \cite{moored_hydrodynamic_2012}, who found optimal efficiencies in both wake types.

\subsection{Modal Decomposition}

\begin{figure}[t!]
\begin{center}
\includegraphics[width=0.6\textwidth]{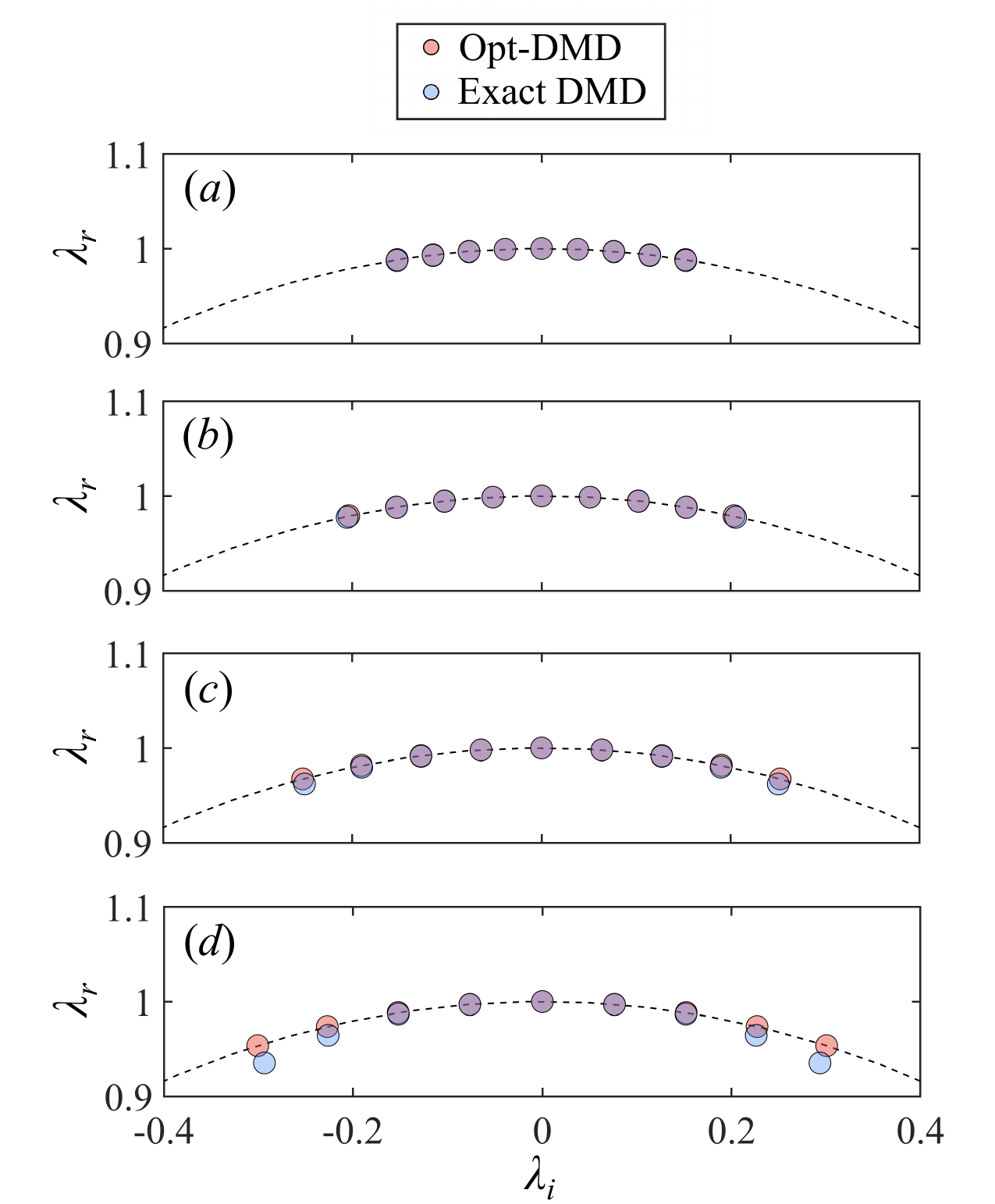}
\caption{
Eigenvalue spectrum for ($a$) $St=0.16$, ($b$) $St=0.23$, ($c$) $St=0.29$, ($d$) $St=0.35$. Purple markers indicate overlap of the opt-DMD (red) and DMD (blue) eigenvalues. $\lambda_r$ denotes the real component of the eigenvalues while $\lambda_i$ denotes the imaginary component.}
\label{fig:opteigs}
\end{center}
\end{figure}

\begin{figure}[t!]
\begin{center}
\includegraphics[width=1\textwidth]{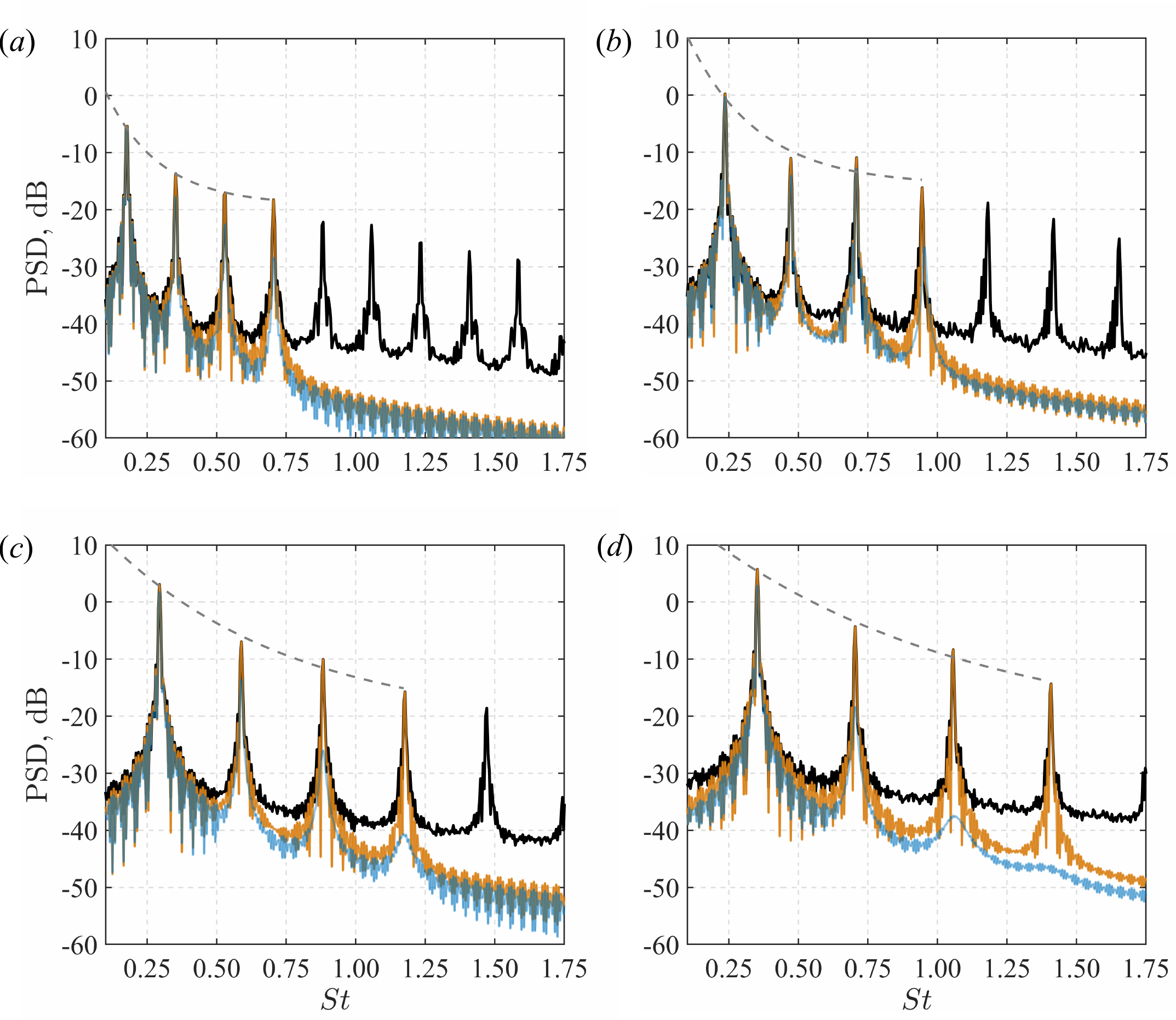}
\caption{Power spectral densities for the  original flow field (black) and the reconstructed flow fields from the opt-DMD (orange) and exact-DMD (blue) method for ($a$) $St=0.16$, ($b$) $St=0.23$, ($c$) $St=0.29$, and ($d$) $St=0.35$. The dotted lines represent exponential decay rates obtained from fits to the first four mode amplitudes from opt-DMD and show the following behavior: $e^{-5.1St}$ ($a$), $e^{-4.0St}$ ($b$), $e^{-1.5St}$ ($c$), and $ e^{-1.0St}$ ($d$).}
\label{fig:psd}
\end{center}
\end{figure}

We now compare results between the exact-DMD and opt-DMD algorithms. Eigenvalues obtained for the first 9 DMD modes for both methods are shown in figure \ref{fig:opteigs}. The mode corresponding to $\lambda = 0$ for each Strouhal number represents the mean flow field, while the other eigenvalues, having a non-zero imaginary part, represent the time-periodic modes. Note that the eigenvalues with non-zero imaginary components are represented in complex conjugate pairs (i.e., with $\lambda_k = \lambda_{k,r} \pm \lambda_{k,i}$), which together represent a single frequency component of the flow field.

As shown by \cite{magionesi_modal_2018}, the eigenvalue problem degenerates and results in harmonic solutions when the dominant components of the flow travel at a constant speed, and with constant shape. This is noticeable in both the opt-DMD and exact-DMD modes, whereby the eigenvalues have oscillation frequencies ($|\omega_{k,i}|$) that are multiples of the first harmonic. The opt-DMD eigenvalues for all Strouhal numbers lie directly on the unit circle, suggesting that the modes neither grow or decay overtime ($|\lambda_k|=1$). In contrast, the exact DMD eigenvalues tend to fall within the circle for increasing $St$, particularly for large values of $|\lambda_{k,i}|$, signifying that the oscillatory modes obtained via exact DMD decay over time ($|\lambda_k|<1$). This is most evident for the largest Strouhal number case ($St=0.35$) in figure \ref{fig:opteigs}$d$, which shows eigenvalues within the unit circle for $|\lambda_{k,i}| > 0.2$. Similar trends were reported by \cite{bagheri_effects_2014}, who observed that eigenvalue damping increases linearly with noise amplitude and quadratically with frequency. For the remainder of this paper, we refer to the flow field produced by the dynamic modes associated with a complex conjugate pair of eigenvalues as a \textit{single mode}, and consider the first to fourth harmonics as \textit{modes} one to four.

To further understand the effects of mode decay via eigenvalue damping, we computed the periodogram-based power spectral density of the original flow field and the reconstructed flow from opt-DMD and exact-DMD modes (see figure \ref{fig:psd}). Note that each of the peaks represents the flow field produced by a \textit{pair} of dynamic modes (i.e., corresponding to a complex conjugate pair of eigenvalues). The normalized frequencies for these peaks, $St= f A_{TE}/U_{\infty}$, represent harmonics of the foil oscillation frequency. The power from the exact-DMD reconstruction are lower in power compared to the opt-DMD reconstruction and full flow field (figure~\ref{fig:psd}$c$ and $d$). This decay can be attributed to the dampened eigenvalues from the exact-DMD algorithm, observed in figure~\ref{fig:opteigs}($c$,$d$). As a result, the effect of reduced power worsens with increasing $St$. These results suggest that the opt-DMD algorithm outperforms the exact-DMD method in retaining the energy in all modes, as also shown by \cite{strom_near-wake_2022}. 
 
 For completeness, we also highlight the exponential decay of the amplitude peaks of the opt-DMD modes, as shown in the dotted line of figure. Interestingly, the exponential fits for $St = 0.16$ and $St = 0.23$ show some flattening at the higher frequencies, suggesting that the wake is not strongly locked to the actuation frequency for these cases. For example, in the case of $St=0.23$ (figure \ref{fig:psd}$b$), the third opt-DMD and exact-DMD modes have a slightly higher energy ($-10.9$ dB and $-12.9$ dB) compared to mode two ($-11.1$ dB and $-14.0$ dB).  The other cases do not show this behavior. 

\subsubsection{Opt-DMD Coherent Structures}

\begin{figure}[t!]
\begin{center}
\includegraphics[width=0.85\textwidth]{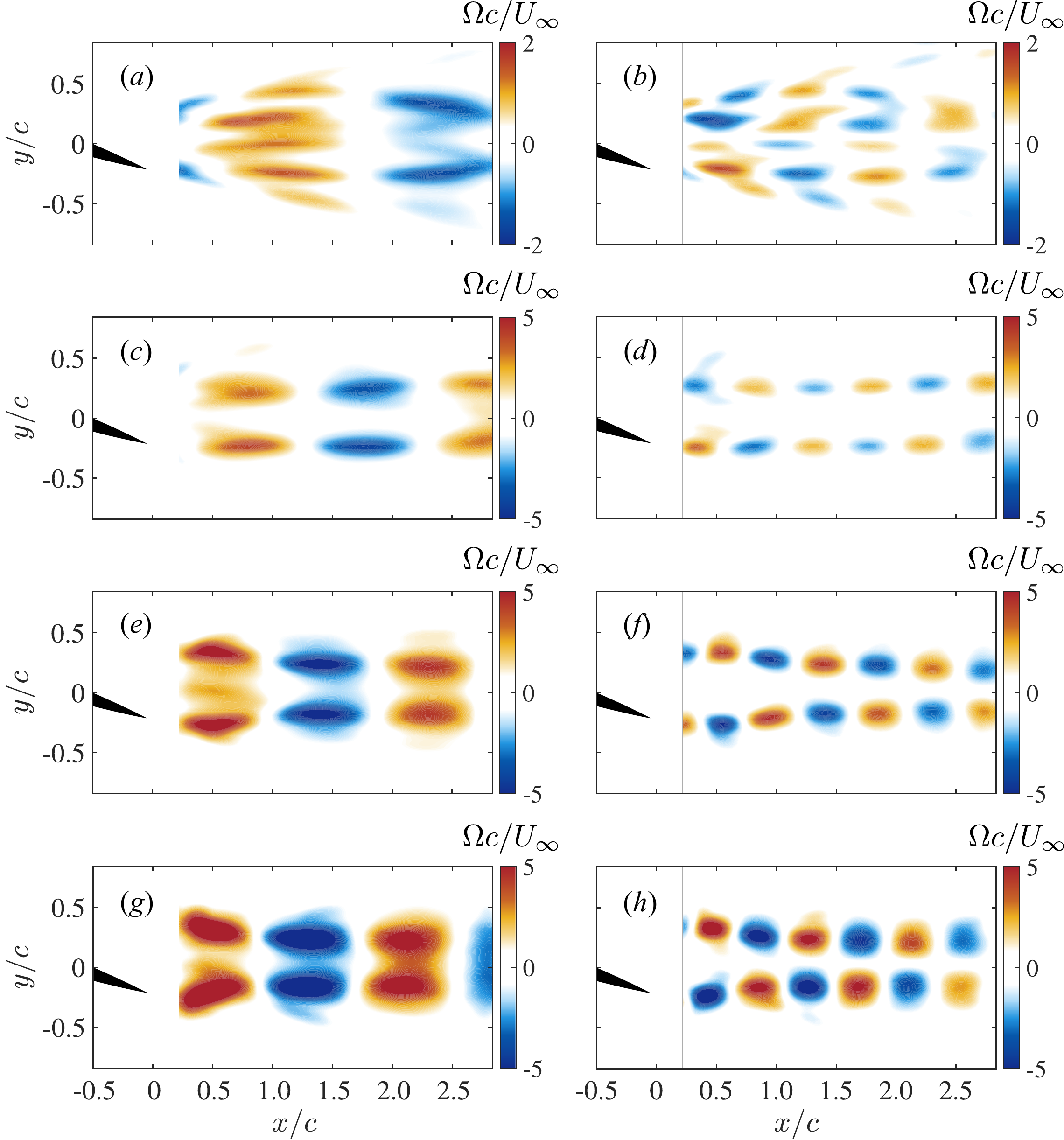}
\caption{Vorticity fields from opt-DMD modes 1 (left) and 2 (right) for ($a,b$) $St=0.16$, ($c,d$) $St=0.23$, ($e,f$) $St=0.29$, and ($g,h$) $St=0.35$. See also supplementary movie 2 and movie 3}
\label{fig:TempVortMode12}
\end{center}
\end{figure}

\begin{figure}[t!]
\begin{center}
\includegraphics[width=0.85\textwidth]{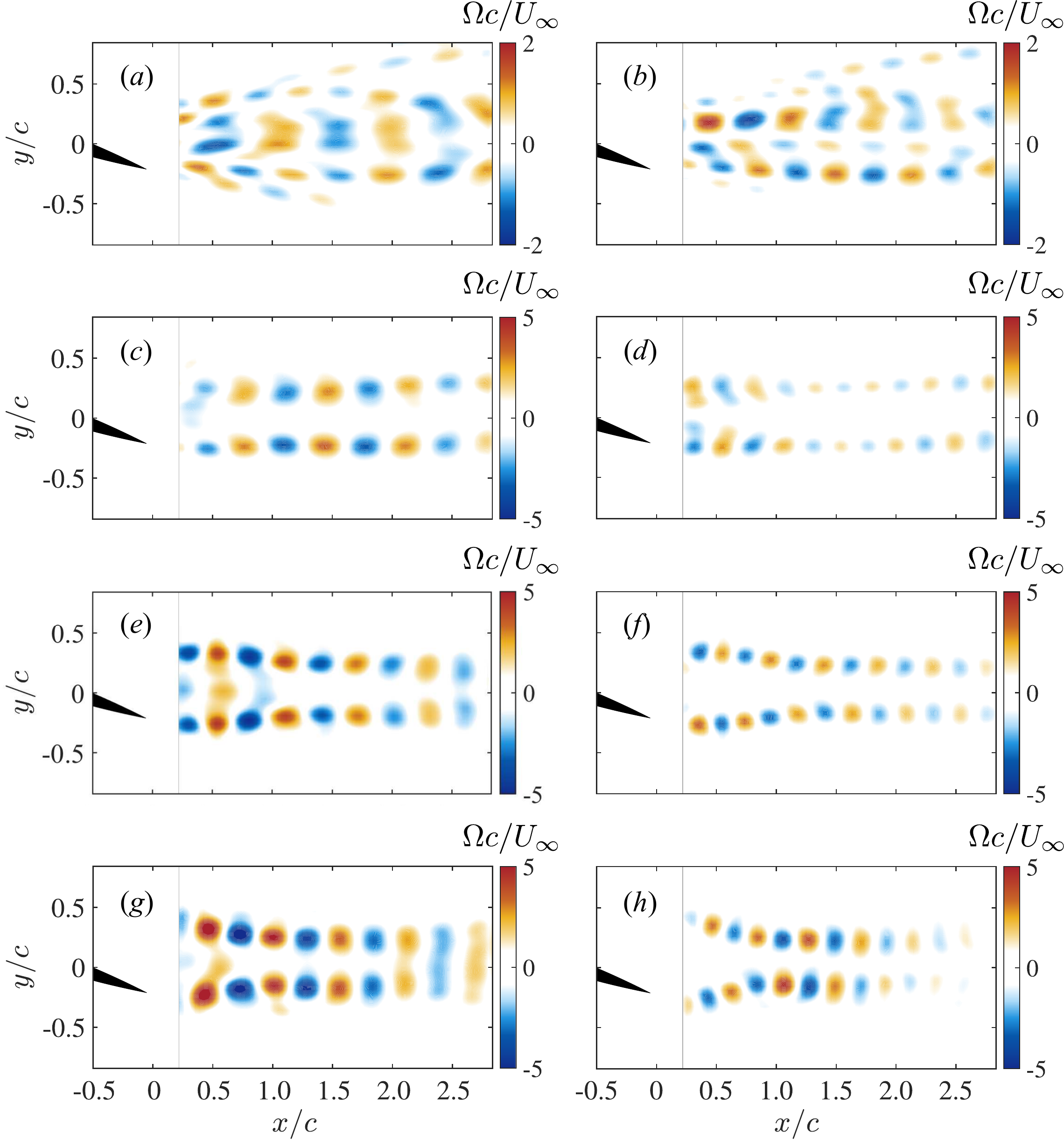}
\caption{Vorticity fields from opt-DMD modes 3 (left) and 4 (right) for ($a,b$) $St=0.16$, ($c,d$) $St=0.23$, ($e,f$) $St=0.29$, and ($g,h$) $St=0.35$. See also supplementary movie 4 and  movie 5.}
\label{fig:TempVortMode34}
\end{center}
\end{figure}

In addition to characterizing oscillation frequencies and growth or decay rates, DMD also provides insight into wake structure. Reconstructions for the first two modes are shown in figure \ref{fig:TempVortMode12}. Each modal reconstruction corresponds to the phase of oscillation shown in figure \ref{fig:VortPIV}. For all $St$ values, mode 1 exhibits top-bottom symmetry in vortex structure across the centerline ($y/c = 0$). In contrast, mode 2  has an antisymmetric structure with vortex lobes of smaller size compared to mode 1. The higher order modes 3 and 4 are also symmetric and asymmetric respectively, as shown in figure \ref{fig:TempVortMode34}. Qualitatively, these modes better capture skews in the wake structure across the centerline, particularly the for the case of $St=0.16$ (figures \ref{fig:TempVortMode34}$a$--$b$).

It was shown by \cite{moored_hydrodynamic_2012} that the overlap in vorticity modes at certain locations in space contributes to producing the full vortex structure. 
Similar observations can also be made in the present study, where the modal hierarchy is trying to represent a convective process in terms of purely oscillatory DMD modes that are harmonics of the flapping frequency. This is reflected in the DMD mode structure: since mode 2 exhibits twice the oscillation frequency as mode 1, vortex lobes from mode 2 exhibit roughly twice the spatial wavenumber as mode 1 (see figure~\ref{fig:TempVortMode12}). A coherent vortex is formed at a particular region in space when both modes are either in-phase or in anti-phase and vortex lobes with the same sign overlaying each other. This is also consistent with the fact that the foil sheds a vortex each half cycle. Whether the vortex formed is positive or negative is determined by the sign of the spatial eigenfunctions. These features (and the modes themselves) are phase-locked in a reference frame moving with the local convective velocity.

 Notable differences between the modes from the near-momentumless wake and that from the 2P and 2S wakes are observed. For instance, mode 1 for $St=0.16$ has approximately four symmetric lobes of vorticity along the lateral axis (figure \ref{fig:TempVortMode12}$a$). These elongated vortex packets expand laterally as they advect downstream and similar trends can be seen for mode 2 (figure \ref{fig:TempVortMode12}$b$). This is consistent with the lateral variation observed in the mean velocity profile and phase-averaged vorticity fields. In contrast, mode 1 for the cases $St=0.23$, $0.29$, and $0.35$ primarily consists of two vortex lobes with decreasing streamwise extent. Thus, the size of the lobes reflects the length scale of the shed vortices, which is partially determined by the convective length scale $U_\infty/f$. 

The differences that reflect the transition from the 2P wake to the 2S reverse von K\'{a}rm\'{a}n street are more subtle. In the case of $St=0.23$ which exhibits a 2P wake morphology, the vorticity lobes travel consistently downstream (see figures \ref{fig:TempVortMode12}$c,d$). As the Strouhal number increases from $St=0.23$ to $St=0.35$, the lobe pairs move inward towards the centerline over the region $0.22 \le x/c \le 1.0$. In the region where $x/c>1.0$, the lobes consistently travel in the streamwise direction again. This dynamic characteristic can be observed better from modes 3 and 4 in figure \ref{fig:TempVortMode34}($c-h$), where the lobes structures are considerably smaller. 
Another key distinction across the 2P to 2S wake transition is that the symmetric vorticity lobes of modes 1 and 3 start to coalesce into a single lobe towards the downstream end of the field of view, $x/c \geq 2.0$. The antisymmetric pairs for modes 2 and 4 remain apart. This is most noticeable for the case of $St=0.35$ (figures \ref{fig:TempVortMode12}$g$ and \ref{fig:TempVortMode34}$g$).

Although the high-frequency wake structures exhibit reverse von K\'{a}rm\'{a}n streets --- as expected for thrust-producing systems --- their symmetric and antisymmetric topologies are similar to that from the classical K\'{a}rm\'{a}n vortex street shed from two-dimensional objects \citep{tu_dynamic_2014, taira_modal_2020, araya_transition_2017}. In these drag-producing wakes, the dominant DMD or POD modes obtained for these flows can also correspond to a series of frequency harmonics. One distinction from the classical drag-producing bluff body wakes is that their symmetric mode structures may consist of a single lobe of vorticity across the $y$-axis --- as shown in experiments from \citep{tu_dynamic_2014} --- rather than two or more from our study. 

\subsubsection{Influence of Modes on Wake Dynamics}

\begin{figure}[t!]
\begin{center}
\includegraphics[width=0.85\textwidth]{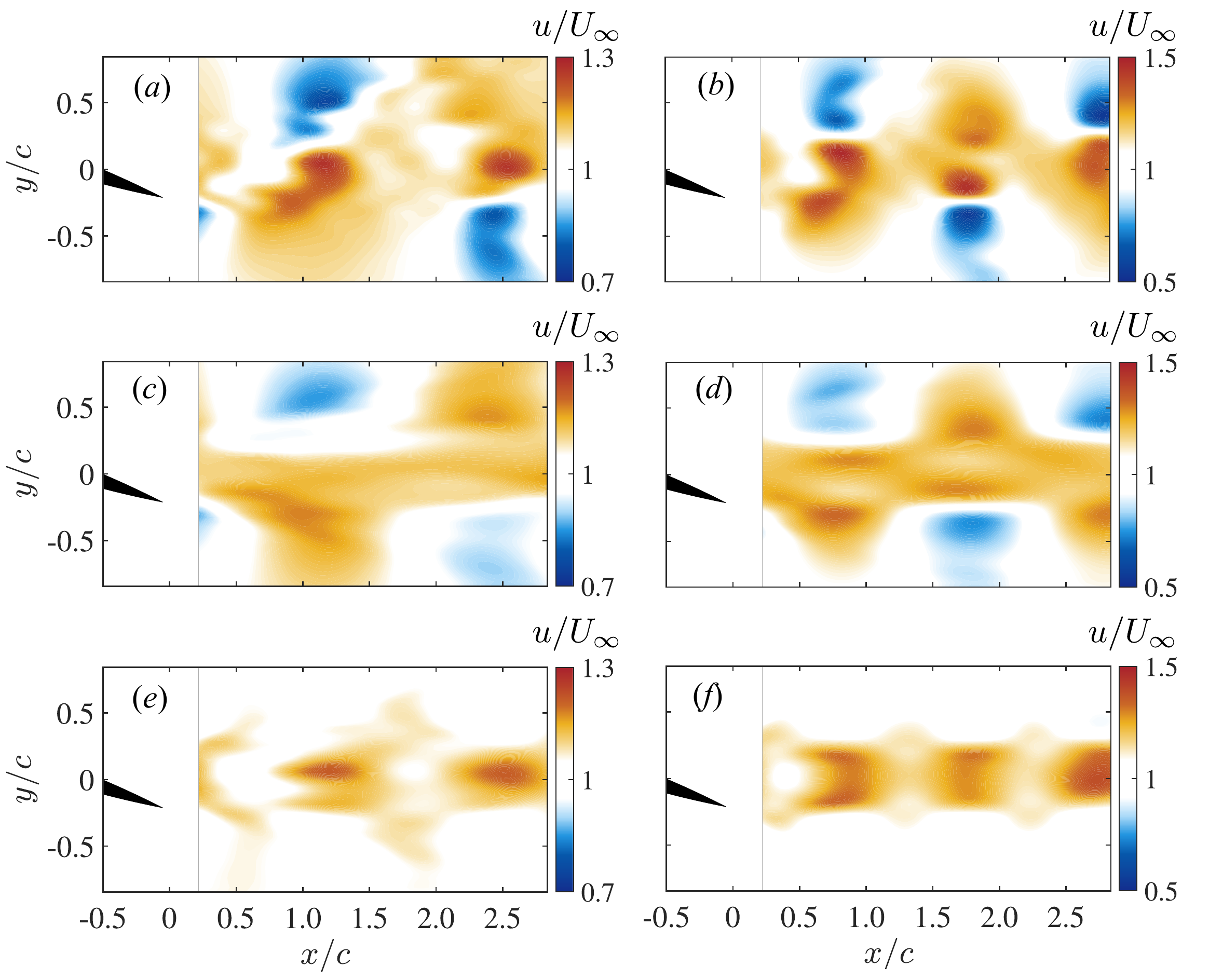}
\caption{Reconstructed streamwise velocity fields: $u = \bar{u} + \sum_{n=1}^{4} \tilde{u}_n$ ($a,b$); $\bar{u} + \tilde{u}_1$ ($c,d$); and $\bar{u} + \tilde{u}_2$ ($e,f$). The left and right columns are of the Strouhal numbers $St=0.16$ and $St=0.23$ respectfully. The mode hierarchy is in terms of frequency, with the lowest corresponding to mode 1.}
\label{fig:modemean1}
\end{center}
\end{figure}

\begin{figure}[t!]
\begin{center}
\includegraphics[width=0.85\textwidth]{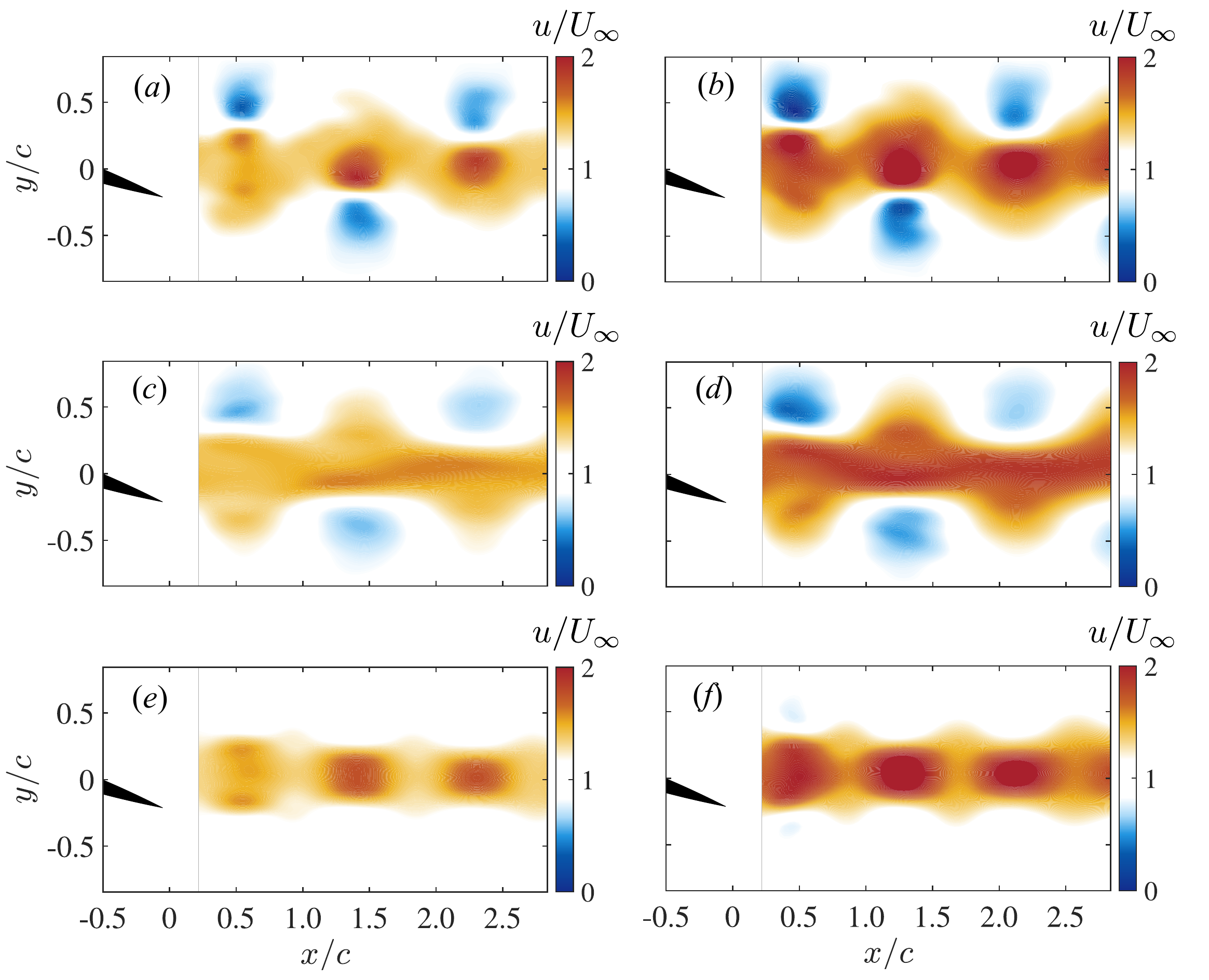}
\caption{Reconstructed streamwise velocity fields: $u = \bar{u} + \sum_{n=1}^{4} \tilde{u}_n$ ($a,b$); $\bar{u} + \tilde{u}_1$ ($c,d$); and $\bar{u} + \tilde{u}_2$ ($e,f$). The left and right columns are of the Strouhal numbers $St=0.29$ and $St=0.35$ respectfully. The mode hierarchy is the same as in figure \ref{fig:modemean1}.}
\label{fig:modemean2}
\end{center}
\end{figure}

To further understand how the symmetric and antisymmetric modes relate to the full flow field, we superimpose the streamwise component associated with DMD modes 1 ($\tilde{u}_1$) and 2 ($\tilde{u}_2$) separately with the time-averaged streamwise velocity ($\bar{u}$). Figures \ref{fig:modemean1} and \ref{fig:modemean2} show the full opt-DMD reconstructions (i.e., comprising modes 1-4) as well as these single-mode reconstructions of the streamwise velocity component. In the original streamwise fields, the vortex development is characterized by the regions where $u$ is less than the that of the freestream velocity ($u/U_\infty<1$), resulting in shear layer roll up. In the transition to the 2S wake, a wavy jet is observed in the original flow fields (\ref{fig:modemean2}$a,b$). Mode 1 with the mean flow ($u = \bar{u} + \tilde{u}_1$) reproduces most of these wake dynamics, including a reasonable portion of the shear layer roll up from which the vortices emerge. 

The wake dynamics associated with mode 1 and the associated symmetric vorticity perturbations closely resemble the spatial instabilities observed by \cite{moored_hydrodynamic_2012} in the 2S wake structure created by a flexible fin. The main distinction lies in the number of vorticity lobes, with the former exhibiting two lobes instead of three. This observation suggests that the opt-DMD modes yield coherent structures that are related to spatial instabilities. It is also likely that mode 1 induces a majority of the net thrust since this mode is associated with the lower velocity structures that form the vortex structures outside the central jet. In contrast, mode 2 is associated with the shedding frequency of each vortex or vortex pair for the 2S and 2P wakes (i.e., twice the oscillation frequency). Mode 2 accounts for much of the remaining shear in the center jet region; this is particularly evident in the higher Strouhal number wakes (figure \ref{fig:modemean2}$e-f$). 

\subsubsection{Coherent Reynolds Stress Contributions from DMD Modes}

The subtle differences in DMD mode structure across Strouhal numbers suggest that the associated Reynolds shear stresses may provide additional insight into thrust and drag effects. Following \cite{reynolds_mechanics_1972} and assuming that the Reynolds stress contributions from the turbulent fluctuations $\mathbf{u}'$ are negligible compared to the contributions from the phase-averaged periodic components $\mathbf{\tilde{u}}$, the time-averaged momentum equation in the streamwise direction can be expressed as follows:
\begin{equation} \label{eq:x-momentum}
\bar{u} \frac{\partial \bar{u}}{\partial x} + \bar{v} \frac{\partial \bar{u}}{\partial y} \approx \frac{\partial}{\partial y} \left[ \nu \left( \frac{\partial \bar{u}}{\partial y} \right) - \overline{\tilde{u} \tilde{v}} \right].
\end{equation}
The equation above also assumes no external pressure gradient, a purely 2D flow, and that cross-stream gradients of the viscous and Reynolds stress terms dominate over the streamwise gradients. Invoking both a strong parallel flow assumption, whereby $\bar{v} \approx 0$ and $\partial \bar{u}/\partial x \approx 0$ locally, equation \ref{eq:x-momentum} can be further simplified to:
\begin{equation} \label{eq:uv}
    \nu\frac{d^2\bar{u}_n}{dy^2}\approx\frac{d\overline{\tilde{u}_n\tilde{v}_n}}{dy},
\end{equation} 
which can be used to \textit{estimate} the induced mean flow. 
In the above equation, $\bar{u}_n$ can be thought of as the mean velocity induced by DMD mode $n$, i.e., the Reynolds shear stresses generated by the periodic velocity components $\tilde{u}_n$ and $\tilde{v}_n$ associated with mode $n$. Since the DMD modes are harmonics of the foil oscillation frequency, there are no mean Reynolds shear stress contributions from interactions across modes.  In other words, we expect $\overline{\tilde{u}_m \tilde{v}_n} = 0$ for $m\neq n$. Equation \eqref{eq:uv} indicates that positive values of $\frac{d\overline{\tilde{u}_n\tilde{v}_n}}{dy}$ are associated with minima in $\bar{u}_n$ and vice versa. Positive $\bar{u}_n$ contributions increase momentum in the wake and produce thrust. The opposite is true for negative $\bar{u}_n$ contributions, which result in induced drag.

Predictions made using the highly simplified momentum balance in equation \ref{eq:uv} are complementary to the control volume analyses of momentum entrainment and expulsion presented by \cite{moored_linear_2014}, which showed that the surrounding fluid is entrained into the near wake region close the foil at wake resonance. The increase in mass flow rate of the jet wake produces thrust, and the entrainment region should therefore have a Reynolds stress distribution in which the periodic components $\overline{\tilde{u}\tilde{v}}$ induce a jet-like mean profile. We solve equation \ref{eq:uv} numerically for $\bar{u}_n$ using the Reynolds shear stress profiles for DMD modes 1 and 2 that are closest to the foil ($x/c=0.22$), enforcing the boundary conditions $\bar{u}_n(-\infty)=\bar{u}_n(\infty)=0$. 

\begin{figure}[t!]
\begin{center}
\includegraphics[width=\textwidth]{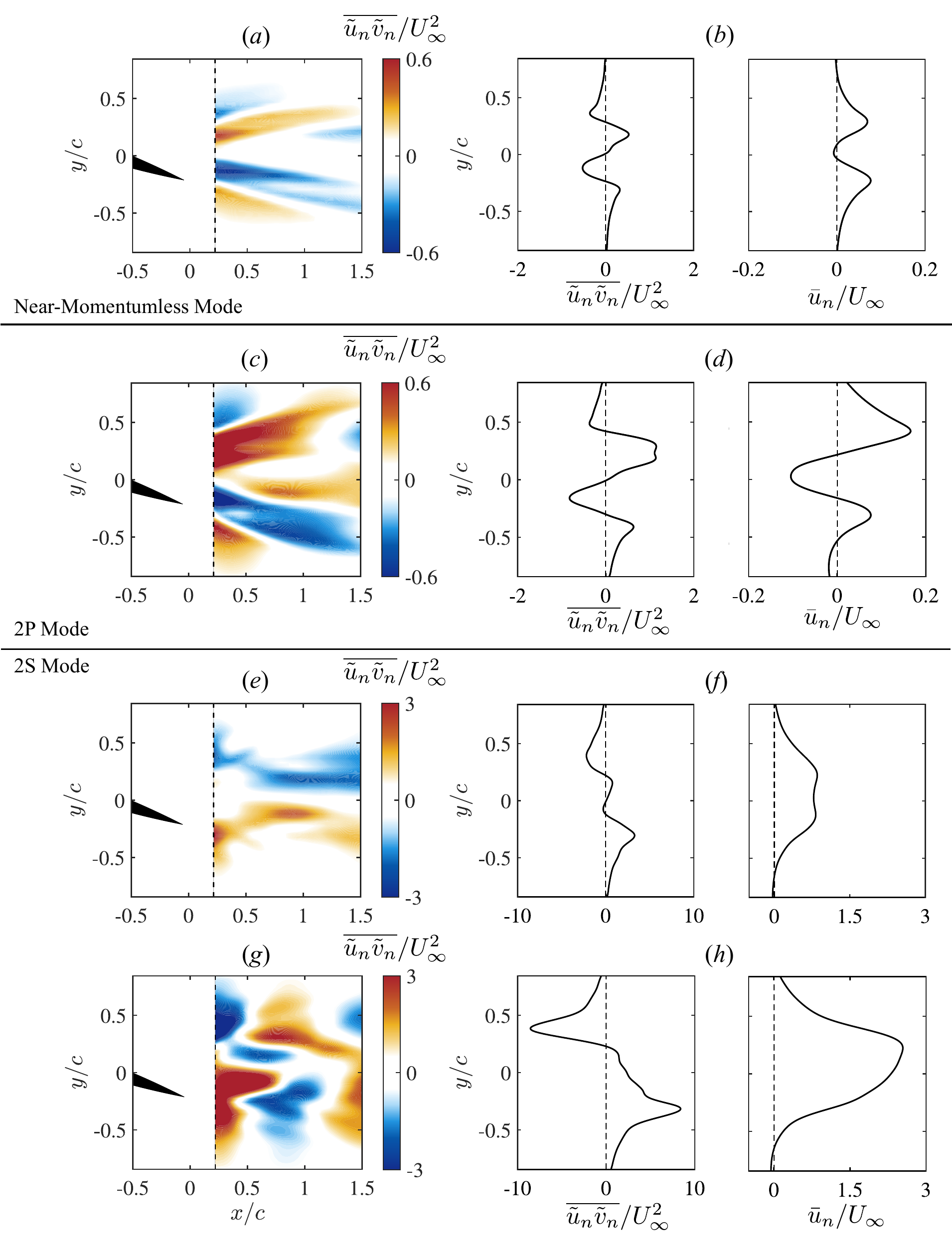}
\caption{Mean Reynolds stress fields $\overline{\tilde{u}_1\tilde{v}_1}$ and induced mean flow profiles $\bar{u}_1$ ($x/c=0.22$) for opt-DMD mode 1 at ($a,b$) $St=0.16$, ($c,d$) $St=0.23$, ($e,f$) $St=0.29$, and ($g,h$) $St=0.35$.} 
\label{fig:Rstress1}
\end{center}
\end{figure}

\begin{figure}[t!]
\begin{center}
\includegraphics[width=\textwidth]{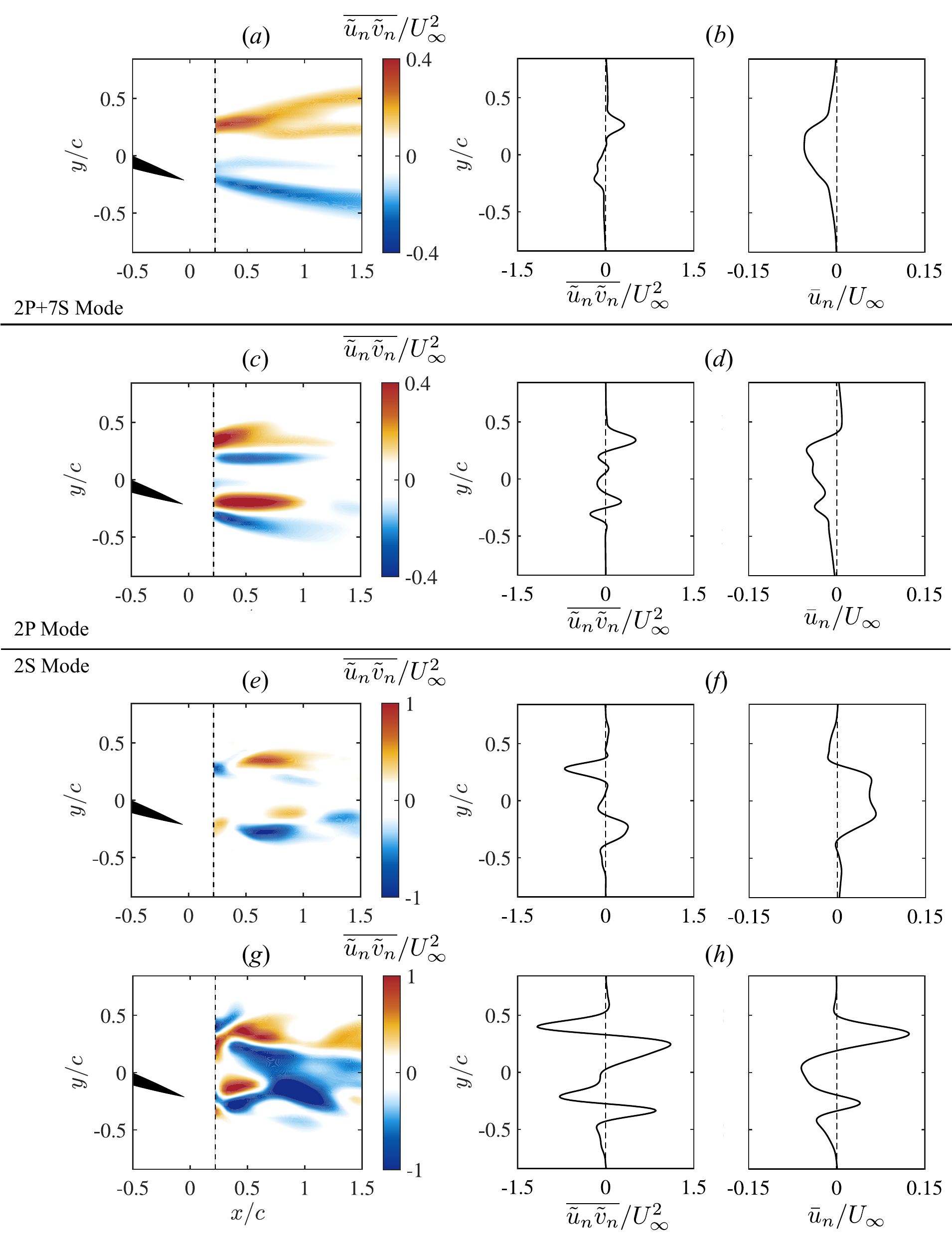}
\caption{Mean Reynolds stress fields $\overline{\tilde{u}_2\tilde{v}_2}$ and induced mean flow profiles $\bar{u}_2$ ($x/c=0.22$) for opt-DMD mode 2 at ($a,b$) $St=0.16$, ($c,d$) $St=0.23$, ($e,f$) $St=0.29$, and ($g,h$) $St=0.35$.}
\label{fig:Rstress2}
\end{center}
\end{figure}

Figures \ref{fig:Rstress1} and \ref{fig:Rstress2} respectively show the coherent Reynolds stress fields $\overline{\tilde{u}_n\tilde{v}_n}$ and induced mean profiles $\bar{u}_n$ predicted using equation (\ref{eq:uv}) for opt-DMD modes 1 and 2. The Reynolds stress fields show patterns that are antisymmetric across the centerline over the region of $0.22 < x/c < 1.0$. This antisymmetry begins to break down for the case of $St=0.35$ (figure \ref{fig:Rstress1}$g$ and \ref{fig:Rstress2}$g$). As expected, mode 1 (i.e., the primary harmonic) generates larger Reynolds shear stresses and induced mean flow contributions compared to mode 2. However, this discrepancy is more pronounced for the higher Strouhal number cases ($St = 0.29, 0.35$) exhibiting 2S-type wakes. The maximum magnitudes of $\overline{\tilde{u}_1\tilde{v}_1}$ for mode 1 increase with Strouhal number. 

The shape of the induced mean flow profiles for mode 1 are similar for the $St=0.16$ and $0.23$ cases (figure \ref{fig:Rstress1}$b,d$), where two distinct positive maxima can be seen (with one noticeable negative component for the case $St=0.23$). Induced mean profiles for both 2S wakes ($St=0.29, 0.35$) in \ref{fig:Rstress1}($f,h$) also show similar characteristics. However, $\bar{u}_1$ for $St = 0.29$ shows two separate maxima while the profile for $St = 0.35$ shows a single maximum near the centerline. This transition from two distinct maxima in the mean profile to a single peak is consistent with the 2P-2S transition noted earlier as the Strouhal number increases from $St = 0.23$ to $St = 0.35$. Interestingly, $\bar{u}_1$ profiles for $St = 0.29$ and $St = 0.35$ match the measured mean profile shapes shown in figure \ref{fig:VortPIV}($c,d$). Specifically, the $St = 0.29$ profile is approximately trapezoidal in shape while the $St = 0.35$ profile resembles a typical jet. Induced $\bar{u}_1$ profiles for these high Strouhal number cases are also much larger in magnitude compared to the lower Strouhal number cases ($St=0.16, 0.23$), which is indicative of higher entrainment and thrust production.

Generally, Reynolds shear stress contributions and induced velocities from mode 2 in figure \ref{fig:Rstress2} are lower than those from mode 1. Particularly, $\bar{u}_2 \ll \bar{u}_1$ for the 2S wakes in $St = 0.29$ and $St = 0.35$. Additionally, While the mode 1 contributions vary significantly in magnitude across Strouhal number, all of the mode 2 profiles exhibit similar maximum values of $\overline{\tilde{u}_2\tilde{v}_2}$ and $\bar{u}_2$. Reynolds stress fields associated with mode 2 for $St=0.16$ and $St = 0.23$ (figure \ref{fig:Rstress2}$b,d$) remain reasonably coherent over the field of view. The resulting mean flow profiles also show consistent negative values ($\bar{u}_2 < 0$), which is indicative of drag generation. This aligns with \cite{Floryan}, who found that scaling laws differed from experiments in efficiency due to viscous drag effects at lower Strouhal numbers. the coherent stress fields are likely characteristics of the transition to bluff body shedding. Note that these drag-inducing effects are characterized by modes with a higher frequency than that from typical bluff body shedding \citep[c.f., similar observations from][]{strom_near-wake_2022}. 

While the $\bar{u}_2$ profiles are primarily negative for the lower Strouhal number cases, distinct positive regions are observed as the flow transitions from the 2P wake to a 2S wake for $St=0.29$ and $0.35$ as seen in figure \ref{fig:Rstress2}($f,h$). For $St = 0.29$, the $\bar{u}_2$ profile is positive and approximately trapezoidal in shape, matching the mean flow in figure \ref{fig:VortPIV}$f$. For $St = 0.35$, the induced $\bar{u}_2$ profile has a mix of positive and negative regions, with the total area under the profile yielding a net negative value. The transition from purely negative $\bar{u}_2$ contributions for $St = 0.16$ and $0.23$ to primarily positive $\bar{u}_2$ contributions for $St=0.29$ is reminiscent of the earlier convective instability work --- as noted by \cite{triantafyllou_optimal_1993}, in a reverse von K\'{a}rm\'{a}n vortex street, there is little to no competition between the drag wake and the thrust wake. This is consistent with the present observations for $St=0.29$.

\subsection{Deterioration of Propulsive Efficiency at High $St$}

\begin{figure}[t!]
\begin{center}
\includegraphics[width=0.75\textwidth]{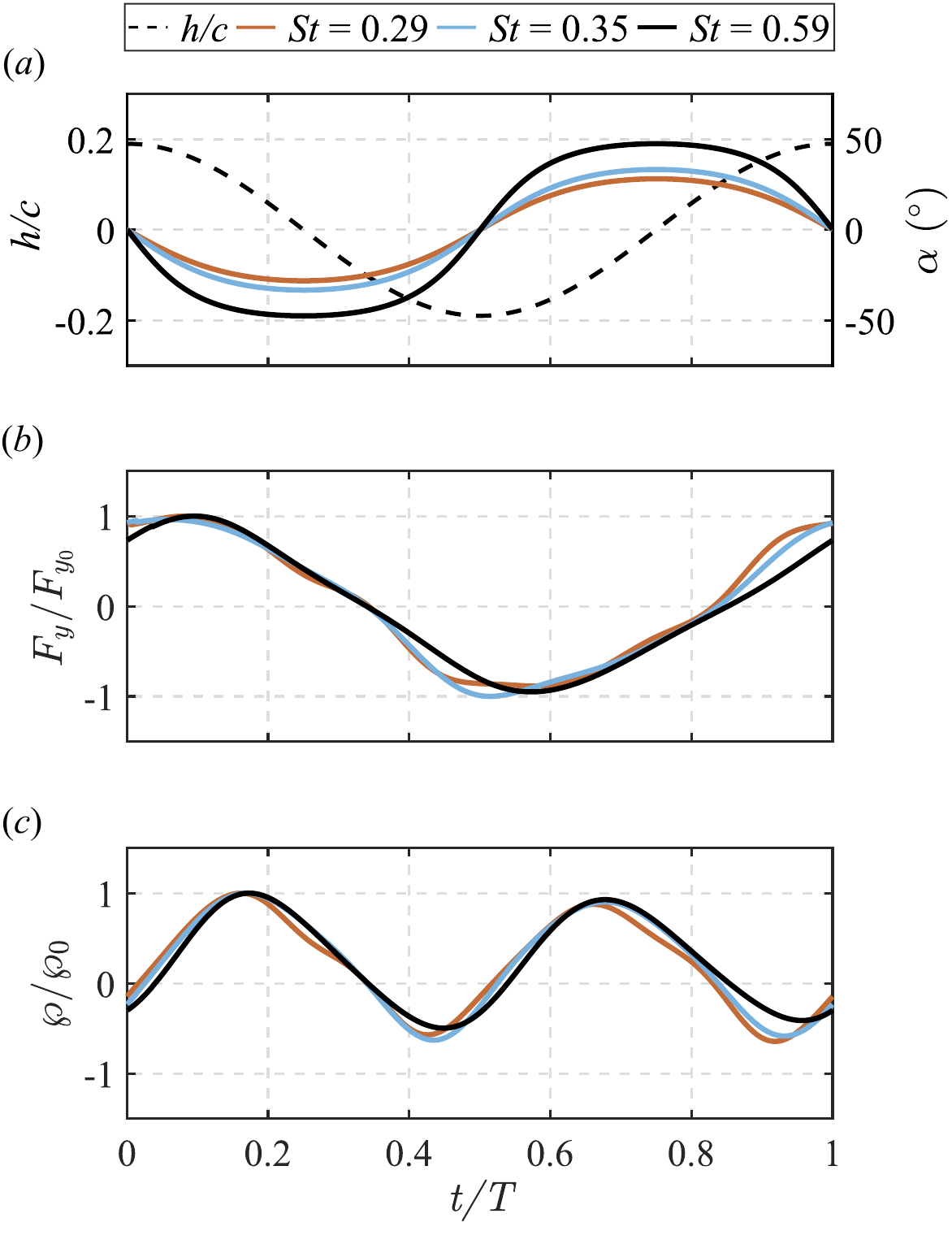}
\caption{Dimensionless heave cycle $h/c$ (dotted line) and effective angle of attack $\alpha$ (solid lines) for one cycle ($a$), together with phase-averaged lift forces $F_y$ ($b$) and power requirements $\wp$ ($c$) normalized by maximum values ($F_{y_0}$ and $\wp_0$) for $St=0.29$ (red), $St=0.35$ (blue), and $St=0.59$ (black).}
\label{fig:phaseavgsig}
\end{center}
\end{figure}

The marginal deterioration in propulsive efficiency along with the negative $\bar{u}_2$ contributions for the $St = 0.35$ case can potentially be attributed to effects from leading-edge vortices. While we do not have access to PIV data around the foil, force and torque measurements, along with the angle of attack trends support this interpretation. In particular, we expect these effects to be evident in the normal or lift forces on the foil, which also influence the power input and propulsive efficiency (see equations~\ref{eq:CP_CT} and \ref{eq:eta}). Dimensionless phase-averaged lift forces $F_y$ and power requirements $\wp$ measured for Strouhal numbers $St = 0.29-0.59$ are shown in figure~\ref{fig:phaseavgsig}. As expected, the lift forces exhibit the same oscillation frequency as the actuation, which also matches the first mode frequencies. The relative power exhibits oscillations at twice the actuation frequency, which is also the second mode frequency. 

Noticeable deviations in relative lift and power are observed between the near-optimal Strouhal numbers ($St=0.29$ and $0.35$) and the highest Strouhal number case ($St=0.59$). Particularly, the small enhancements in the magnitude of the lift force observed between $t/T\approx 0.4-0.5$ and $t/T \approx 0.9-1.0$ lead to lower relative power requirements, as shown in figures~\ref{fig:phaseavgsig}($b,c$). For the highest propulsive efficiency case of $St=0.29$, additional decreases in relative power at $t/T \approx 0.25$ and $0.75$ lower the overall power across the full oscillation cycle (figure \ref{fig:phaseavgsig}$c$). Note that the relative lift force enhancements for this Strouhal number case appear soon after maxima and minima in the effective angle of attack ($t/T = 0.25, 0.75$). 


These relative lift force enhancements may be linked to the structure of the Reynolds shear stress contribution and positive induced velocity from DMD mode 2 for $St=0.29$ (figures~\ref{fig:Rstress2}$e,f$), suggesting a delay in separation of the leading edge vortices. The separation effects may start to occur for the $St = 0.35$ case, which show a negative induced velocity from mode 2 (figure \ref{fig:Rstress1}$h$) and higher relative power requirements in comparison to those from $St=0.29$, resulting in lower propulsive efficiency. The deterioration in performance is even more visible for the $St=0.59$ case, which shows a substantial decrease in the magnitude of $F_y/F_{y0}$ and corresponding increases in $\wp/\wp_0$. 

The tapering of efficiency past $St=0.29$ can also be explained based on high angles of attack. The stall angle for a static NACA 0012 foil is approximately $16^\circ$. However, oscillating loads can increase the stall angle substantially \citep{maresca_experiments_1979}. For the present experiments, the maximum angles of attack $\alpha_0$ for the higher Strouhal number cases $St=0.29$, $0.35$, and $0.59$ are $28.2^\circ$, $33.1^\circ$, and $47.5^\circ$, respectively. Thus, it is likely that the induced drag effects from the secondary mode (figure \ref{fig:Rstress2}$h,f$) become more relevant as the effective angle of attack increases, through which separation effects may occur --- as also evident in the $F_y/F_{y0}$ traces. In turn, the efficiency tapers downward past the Strouhal number $St=0.29$. However, it would be misleading to say that all cases with high angles of attack lead to reduced efficiency, as some combinations of kinematic parameters can delay these dynamic stall effects \citep{anderson1998,maresca_experiments_1979,ellington_leading-edge_1996}. Instead, a high angle of attack where the flow stays attached to the foil can potentially maximize efficiency, an effect that has also been observed for flexible propulsors \citep{QuinnOpt}.


\section{Discussion and Conclusions}

A triple decomposition method in which the periodic component of the wake is composed of opt-DMD modes was used to provide further physical insight into the propulsive performance of oscillating foils. The experimental data in which the method was used are broadly consistent with prior literature. We observe a near-momentumless wake structure for $St = 0.16$ and peak propulsive efficiency for $St = 0.29$. Over this range of Strouhal numbers, PIV measurements show a transition from a momentumless wake to a 2S wake morphology associated with the classical reverse von K\'{a}rm\'{a}n vortex street. The opt-DMD method is employed here instead of the classical or (exact) DMD approach as it captures more energy in the periodic component of the flow field. For this particular foil, the opt-DMD modes appear as harmonics of the foil oscillating frequency, with alternating symmetric and antisymmetric morphology across the wake centerline. 
 
The most interesting finding from our study is the relationship between the coherent Reynolds stress contributions from DMD modes and their impact on propulsive performance. Although equation~\ref{eq:uv} is only an approximation to the mean flow equation for the wake, it helps delineate modal contributions to drag and thrust. In our experiments, the mean velocities induced by the primary opt-DMD mode (mode 1) generally show thrust-producing characteristics across all $St$ values, though there is a transition from a two-hump mean profile for the 2P wakes to a jet-like mean profile for the 2S wakes. However, several different trends emerge from the mean velocities induced by mode 2, representing features oscillating at twice the fundamental frequency of the foil: (i) for Strouhal numbers below the optimum ($St<0.29$) the induced mean velocities $\bar{u}_2$ are negative and suggest a transition to bluff body shedding; (ii) the secondary mode for the optimum Strouhal number case ($St=0.29$) has a prominent positive induced mean velocity, indicating that the thrust wake is dominant; (iii) past the optimum Strouhal number, mode 2 transitions back to a negative induced mean velocity, which may be indicative of flow separation as a result of high angles of attack ($St>0.29$). 



Given the close connection between DMD and stability analyses, it may be of interest to compare the opt-DMD modes with features identified in prior stability analyses \citep{moored_hydrodynamic_2012,arbie_stability_2016}. The development of the opt-DMD modes close to the foil suggest that their may be some spatial growth of these perturbations in the near-field. Thus, there may be links between these opt-DMD modes and the spatially-growing modes identified in the `wake resonance' studies \citep{triantafyllou_optimal_1993, lewin_modelling_2003,moored_linear_2014}. Nonetheless, it should be emphasized that the temporal DMD method was used in this study, which extracts modes at a particular oscillation frequency. A spatial DMD approach could provide a more direct approach to understanding these instability mechanisms. In this case, a spatial resolution that is higher than the the PIV data collected in this study would be necessary to allow for an appropriate spatial DMD analysis.


This study focused on a narrow parameter range: we studied a single, rigid foil exhibiting periodic single-frequency oscillations. However, modal analysis methods similar to those employed here could provide substantial insights into wakes involving more complex kinematics, dynamic fluid-structure interactions, multiple-foil interactions, or massively separated flows \citep[see e.g.,][]{raspa_topology-induced_2013,andersen_wake_2017}. 

\section{Acknowledgments}
This material is based on work supported by the National Science Foundation under Grant No. 1943105. The authors thank Prof. Geoffrey Spedding and Steven Brunton for insightful discussions. Declaration of interests: the authors report no conflict of interest.

\bibliography{references}

\begin{thebibliography}{42}
\providecommand{\natexlab}[1]{#1}
\providecommand{\url}[1]{\texttt{#1}}
\expandafter\ifx\csname urlstyle\endcsname\relax
  \providecommand{\doi}[1]{doi: #1}\else
  \providecommand{\doi}{doi: \begingroup \urlstyle{rm}\Url}\fi

\bibitem[Andersen et~al.(2017)Andersen, Bohr, Schnipper, and Walther]{andersen_wake_2017}
A.~Andersen, T.~Bohr, T.~Schnipper, and J.~H. Walther.
\newblock Wake structure and thrust generation of a flapping foil in two-dimensional flow.
\newblock \emph{Journal of Fluid Mechanics}, 812:\penalty0 R4, Feb. 2017.
\newblock ISSN 0022-1120, 1469-7645.
\newblock \doi{10.1017/jfm.2016.808}.
\newblock URL \url{https://www.cambridge.org/core/journals/journal-of-fluid-mechanics/article/wake-structure-and-thrust-generation-of-a-flapping-foil-in-twodimensional-flow/BC3C0588ADE745DD1290EC2FAFF5E6F9}.
\newblock Publisher: Cambridge University Press.

\bibitem[Anderson et~al.(1998)Anderson, Streitlien, Barrett, and Triantafyllou]{anderson1998}
J.~M. Anderson, K.~Streitlien, D.~S. Barrett, and M.~S. Triantafyllou.
\newblock Oscillating foils of high propulsive efficiency.
\newblock \emph{Journal of Fluid Mechanics}, 360:\penalty0 41–72, 1998.
\newblock \doi{10.1017/S0022112097008392}.

\bibitem[Araya et~al.(2017)Araya, Colonius, and Dabiri]{araya_transition_2017}
D.~B. Araya, T.~Colonius, and J.~O. Dabiri.
\newblock Transition to bluff-body dynamics in the wake of vertical-axis wind turbines.
\newblock \emph{Journal of Fluid Mechanics}, 813:\penalty0 346--381, Feb. 2017.
\newblock ISSN 0022-1120, 1469-7645.
\newblock \doi{10.1017/jfm.2016.862}.
\newblock URL \url{https://www.cambridge.org/core/product/identifier/S0022112016008624/type/journal_article}.

\bibitem[Arbie et~al.(2016)Arbie, Ehrenstein, and Eloy]{arbie_stability_2016}
M.~R. Arbie, U.~Ehrenstein, and C.~Eloy.
\newblock Stability of momentumless wakes.
\newblock \emph{Journal of Fluid Mechanics}, 808:\penalty0 316--336, Dec. 2016.
\newblock ISSN 0022-1120, 1469-7645.
\newblock \doi{10.1017/jfm.2016.645}.
\newblock URL \url{https://www.cambridge.org/core/journals/journal-of-fluid-mechanics/article/stability-of-momentumless-wakes/614D5141D37E6A5D123ABC9D11D8B0A7}.
\newblock Publisher: Cambridge University Press.

\bibitem[Askham and Kutz(2018)]{askham_variable_2018}
T.~Askham and J.~N. Kutz.
\newblock Variable {Projection} {Methods} for an {Optimized} {Dynamic} {Mode} {Decomposition}.
\newblock \emph{SIAM Journal on Applied Dynamical Systems}, 17\penalty0 (1):\penalty0 380--416, Jan. 2018.
\newblock \doi{10.1137/M1124176}.
\newblock URL \url{https://epubs.siam.org/doi/abs/10.1137/M1124176}.
\newblock Publisher: Society for Industrial and Applied Mathematics.

\bibitem[Bagheri(2014)]{bagheri_effects_2014}
S.~Bagheri.
\newblock Effects of weak noise on oscillating flows: {Linking} quality factor, {Floquet} modes, and {Koopman} spectrum.
\newblock \emph{Physics of Fluids}, 26\penalty0 (9):\penalty0 094104, Sept. 2014.
\newblock ISSN 1070-6631, 1089-7666.
\newblock \doi{10.1063/1.4895898}.
\newblock URL \url{http://aip.scitation.org/doi/10.1063/1.4895898}.

\bibitem[Baj et~al.(2015)Baj, Bruce, and Buxton]{baj_triple_2015}
P.~Baj, P.~J.~K. Bruce, and O.~R.~H. Buxton.
\newblock The triple decomposition of a fluctuating velocity field in a multiscale flow.
\newblock \emph{Physics of Fluids}, 27\penalty0 (7):\penalty0 075104, July 2015.
\newblock ISSN 1070-6631, 1089-7666.
\newblock \doi{10.1063/1.4923744}.
\newblock URL \url{http://aip.scitation.org/doi/10.1063/1.4923744}.

\bibitem[Beal et~al.(2006)Beal, Hover, Triantafyllou, Liao, and Lauder]{beal_passive_2006}
D.~N. Beal, F.~S. Hover, M.~S. Triantafyllou, J.~C. Liao, and G.~V. Lauder.
\newblock Passive propulsion in vortex wakes.
\newblock \emph{Journal of Fluid Mechanics}, 549\penalty0 (-1):\penalty0 385, Feb. 2006.
\newblock ISSN 0022-1120, 1469-7645.
\newblock \doi{10.1017/S0022112005007925}.
\newblock URL \url{http://www.journals.cambridge.org/abstract_S0022112005007925}.

\bibitem[Bryant et~al.(2012)Bryant, Mahtani, and Garcia]{bryant_wake_2012}
M.~Bryant, R.~L. Mahtani, and E.~Garcia.
\newblock Wake synergies enhance performance in aeroelastic vibration energy harvesting.
\newblock \emph{Journal of Intelligent Material Systems and Structures}, 23\penalty0 (10):\penalty0 1131--1141, July 2012.
\newblock ISSN 1045-389X, 1530-8138.
\newblock \doi{10.1177/1045389X12443599}.
\newblock URL \url{http://journals.sagepub.com/doi/10.1177/1045389X12443599}.

\bibitem[Buren et~al.(2020)Buren, Floryan, and Smits]{buren_floryan_smits_2020}
T.~V. Buren, D.~Floryan, and A.~J. Smits.
\newblock \emph{Bioinspired Underwater Propulsors}, page 113–139.
\newblock Cambridge University Press, 2020.
\newblock \doi{10.1017/9781139058995.006}.

\bibitem[Dawson et~al.(2016)Dawson, Hemati, Williams, and Rowley]{dawson_characterizing_2016}
S.~T.~M. Dawson, M.~S. Hemati, M.~O. Williams, and C.~W. Rowley.
\newblock Characterizing and correcting for the effect of sensor noise in the dynamic mode decomposition.
\newblock \emph{Experiments in Fluids}, 57\penalty0 (3):\penalty0 42, Mar. 2016.
\newblock ISSN 0723-4864, 1432-1114.
\newblock \doi{10.1007/s00348-016-2127-7}.
\newblock URL \url{http://arxiv.org/abs/1507.02264}.
\newblock arXiv:1507.02264 [physics].

\bibitem[Dewey et~al.(2012)Dewey, Carriou, and Smits]{dewey_relationship_2012}
P.~A. Dewey, A.~Carriou, and A.~J. Smits.
\newblock On the relationship between efficiency and wake structure of a batoid-inspired oscillating fin.
\newblock \emph{Journal of Fluid Mechanics}, 691:\penalty0 245--266, Jan. 2012.
\newblock ISSN 1469-7645, 0022-1120.
\newblock \doi{10.1017/jfm.2011.472}.
\newblock URL \url{https://www.cambridge.org/core/journals/journal-of-fluid-mechanics/article/on-the-relationship-between-efficiency-and-wake-structure-of-a-batoidinspired-oscillating-fin/013AD59953E8313AA243A96A85C5A0CB}.
\newblock Publisher: Cambridge University Press.

\bibitem[Ellington et~al.(1996)Ellington, Van Den~Berg, Willmott, and Thomas]{ellington_leading-edge_1996}
C.~P. Ellington, C.~Van Den~Berg, A.~P. Willmott, and A.~L.~R. Thomas.
\newblock Leading-edge vortices in insect flight.
\newblock \emph{Nature}, 384\penalty0 (6610):\penalty0 626--630, Dec. 1996.
\newblock ISSN 0028-0836, 1476-4687.
\newblock \doi{10.1038/384626a0}.
\newblock URL \url{https://www.nature.com/articles/384626a0}.

\bibitem[Eloy(2012)]{eloy_optimal_2012}
C.~Eloy.
\newblock Optimal {Strouhal} number for swimming animals.
\newblock \emph{Journal of Fluids and Structures}, 30:\penalty0 205--218, Apr. 2012.
\newblock ISSN 08899746.
\newblock \doi{10.1016/j.jfluidstructs.2012.02.008}.
\newblock URL \url{http://arxiv.org/abs/1102.0223}.
\newblock arXiv:1102.0223 [physics].

\bibitem[Floryan et~al.(2017)Floryan, Van~Buren, Rowley, and Smits]{Floryan}
D.~Floryan, T.~Van~Buren, C.~W. Rowley, and A.~J. Smits.
\newblock Scaling the propulsive performance of heaving and pitching foils.
\newblock \emph{Journal of Fluid Mechanics}, 822:\penalty0 386–397, Jun 2017.
\newblock ISSN 1469-7645.
\newblock \doi{10.1017/jfm.2017.302}.
\newblock URL \url{http://dx.doi.org/10.1017/jfm.2017.302}.

\bibitem[Floryan et~al.(2020)Floryan, Van~Buren, and Smits]{floryan_swimmers_2020}
D.~Floryan, T.~Van~Buren, and A.~J. Smits.
\newblock Swimmers’ wake structures are not reliable indicators of swimming performance.
\newblock \emph{Bioinspiration \& Biomimetics}, 15\penalty0 (2):\penalty0 024001, Feb. 2020.
\newblock ISSN 1748-3190.
\newblock \doi{10.1088/1748-3190/ab6fb9}.
\newblock URL \url{https://iopscience.iop.org/article/10.1088/1748-3190/ab6fb9}.

\bibitem[Hemati et~al.(2017)Hemati, Rowley, Deem, and Cattafesta]{hemati_-biasing_2017}
M.~S. Hemati, C.~W. Rowley, E.~A. Deem, and L.~N. Cattafesta.
\newblock De-{Biasing} the {Dynamic} {Mode} {Decomposition} for {Applied} {Koopman} {Spectral} {Analysis}.
\newblock \emph{Theoretical and Computational Fluid Dynamics}, 31\penalty0 (4):\penalty0 349--368, Aug. 2017.
\newblock ISSN 0935-4964, 1432-2250.
\newblock \doi{10.1007/s00162-017-0432-2}.
\newblock URL \url{http://arxiv.org/abs/1502.03854}.
\newblock arXiv:1502.03854 [physics].

\bibitem[Hussain and Reynolds(1970)]{hussain_mechanics_1970}
A.~K. M.~F. Hussain and W.~C. Reynolds.
\newblock The mechanics of an organized wave in turbulent shear flow.
\newblock \emph{Journal of Fluid Mechanics}, 41\penalty0 (2):\penalty0 241--258, Apr. 1970.
\newblock ISSN 1469-7645, 0022-1120.
\newblock \doi{10.1017/S0022112070000605}.
\newblock URL \url{https://www.cambridge.org/core/journals/journal-of-fluid-mechanics/article/abs/mechanics-of-an-organized-wave-in-turbulent-shear-flow/8FE76E0E31505D1039E0AB8F249CE304}.
\newblock Publisher: Cambridge University Press.

\bibitem[Jones et~al.(1997)Jones, Platzer, Jones, and Platzer]{jones_numerical_1997}
K.~Jones, M.~Platzer, K.~Jones, and M.~Platzer.
\newblock Numerical computation of flapping-wing propulsion and power extraction.
\newblock In \emph{35th {Aerospace} {Sciences} {Meeting} and {Exhibit}}, Reno,NV,U.S.A., Jan. 1997. American Institute of Aeronautics and Astronautics.
\newblock \doi{10.2514/6.1997-826}.
\newblock URL \url{https://arc.aiaa.org/doi/10.2514/6.1997-826}.

\bibitem[Lentink et~al.(2008)Lentink, Muijres, Donker-Duyvis, and van Leeuwen]{lentink_vortex-wake_2008}
D.~Lentink, F.~T. Muijres, F.~J. Donker-Duyvis, and J.~L. van Leeuwen.
\newblock Vortex-wake interactions of a flapping foil that models animal swimming and flight.
\newblock \emph{Journal of Experimental Biology}, 211\penalty0 (2):\penalty0 267--273, Jan. 2008.
\newblock ISSN 0022-0949.
\newblock \doi{10.1242/jeb.006155}.
\newblock URL \url{https://doi.org/10.1242/jeb.006155}.

\bibitem[Lewin and Haj-Hariri(2003)]{lewin_modelling_2003}
G.~C. Lewin and H.~Haj-Hariri.
\newblock Modelling thrust generation of a two-dimensional heaving airfoil in a viscous flow.
\newblock \emph{Journal of Fluid Mechanics}, 492:\penalty0 339--362, 2003.
\newblock \doi{10.1017/S0022112003005743}.
\newblock Publisher: Cambridge University Press.

\bibitem[Mackowski and Williamson(2015)]{mackowski_williamson_2015}
A.~W. Mackowski and C.~H.~K. Williamson.
\newblock Direct measurement of thrust and efficiency of an airfoil undergoing pure pitching.
\newblock \emph{Journal of Fluid Mechanics}, 765:\penalty0 524–543, 2015.
\newblock \doi{10.1017/jfm.2014.748}.

\bibitem[Magionesi et~al.(2018)Magionesi, Dubbioso, Muscari, and Mascio]{magionesi_modal_2018}
F.~Magionesi, G.~Dubbioso, R.~Muscari, and A.~D. Mascio.
\newblock Modal analysis of the wake past a marine propeller.
\newblock \emph{Journal of Fluid Mechanics}, 855:\penalty0 469--502, Nov. 2018.
\newblock ISSN 0022-1120, 1469-7645.
\newblock \doi{10.1017/jfm.2018.631}.
\newblock URL \url{https://www.cambridge.org/core/journals/journal-of-fluid-mechanics/article/abs/modal-analysis-of-the-wake-past-a-marine-propeller/C27BA0FDB54DE20F6CCC4C7A7CF65F92}.
\newblock Publisher: Cambridge University Press.

\bibitem[Maresca et~al.(1979)Maresca, Favier, and Rebont]{maresca_experiments_1979}
C.~Maresca, D.~Favier, and J.~Rebont.
\newblock Experiments on an aerofoil at high angle of incidence in longitudinal oscillations.
\newblock \emph{Journal of Fluid Mechanics}, 92\penalty0 (4):\penalty0 671--690, June 1979.
\newblock ISSN 1469-7645, 0022-1120.
\newblock \doi{10.1017/S0022112079000823}.
\newblock URL \url{https://www.cambridge.org/core/journals/journal-of-fluid-mechanics/article/experiments-on-an-aerofoil-at-high-angle-of-incidence-in-longitudinal-oscillations/21A7BA70DC7235B790CF2DFC891F753A}.
\newblock Publisher: Cambridge University Press.

\bibitem[McKinney and DeLaurier(1981)]{mckinney_wingmill_1981}
W.~McKinney and J.~DeLaurier.
\newblock Wingmill: {An} {Oscillating}-{Wing} {Windmill}.
\newblock \emph{Journal of Energy}, 5\penalty0 (2):\penalty0 109--115, Mar. 1981.
\newblock ISSN 0146-0412, 1555-5917.
\newblock \doi{10.2514/3.62510}.
\newblock URL \url{https://arc.aiaa.org/doi/10.2514/3.62510}.

\bibitem[Moored et~al.(2012)Moored, Dewey, Smits, and Haj-Hariri]{moored_hydrodynamic_2012}
K.~W. Moored, P.~A. Dewey, A.~J. Smits, and H.~Haj-Hariri.
\newblock Hydrodynamic wake resonance as an underlying principle of efficient unsteady propulsion.
\newblock \emph{Journal of Fluid Mechanics}, 708:\penalty0 329--348, 2012.
\newblock \doi{10.1017/jfm.2012.313}.
\newblock Publisher: Cambridge University Press.

\bibitem[Moored et~al.(2014)Moored, Dewey, Boschitsch, Smits, and Haj-Hariri]{moored_linear_2014}
K.~W. Moored, P.~A. Dewey, B.~M. Boschitsch, A.~J. Smits, and H.~Haj-Hariri.
\newblock Linear instability mechanisms leading to optimally efficient locomotion with flexible propulsors.
\newblock \emph{Physics of Fluids}, 26\penalty0 (4):\penalty0 041905, Apr. 2014.
\newblock ISSN 1070-6631.
\newblock \doi{10.1063/1.4872221}.
\newblock URL \url{https://aip.scitation.org/doi/citedby/10.1063/1.4872221}.
\newblock Publisher: American Institute of Physics.

\bibitem[Quinn et~al.(2015)Quinn, Lauder, and Smits]{QuinnOpt}
D.~B. Quinn, G.~V. Lauder, and A.~J. Smits.
\newblock Maximizing the efficiency of a flexible propulsor using experimental optimization.
\newblock \emph{Journal of Fluid Mechanics}, 767:\penalty0 430–448, 2015.
\newblock \doi{10.1017/jfm.2015.35}.

\bibitem[Raspa et~al.(2013)Raspa, Godoy-Diana, and Thiria]{raspa_topology-induced_2013}
V.~Raspa, R.~Godoy-Diana, and B.~Thiria.
\newblock Topology-induced effect in biomimetic propulsive wakes.
\newblock \emph{Journal of Fluid Mechanics}, 729:\penalty0 377--387, Aug. 2013.
\newblock ISSN 0022-1120, 1469-7645.
\newblock \doi{10.1017/jfm.2013.295}.
\newblock URL \url{https://www.cambridge.org/core/product/identifier/S0022112013002954/type/journal_article}.

\bibitem[Reynolds and Hussain(1972)]{reynolds_mechanics_1972}
W.~C. Reynolds and A.~K. M.~F. Hussain.
\newblock The mechanics of an organized wave in turbulent shear flow. {Part} 3. {Theoretical} models and comparisons with experiments.
\newblock \emph{Journal of Fluid Mechanics}, 54\penalty0 (2):\penalty0 263--288, July 1972.
\newblock ISSN 0022-1120, 1469-7645.
\newblock \doi{10.1017/S0022112072000679}.
\newblock URL \url{https://www.cambridge.org/core/product/identifier/S0022112072000679/type/journal_article}.

\bibitem[Sarmast et~al.(2014)Sarmast, Dadfar, Mikkelsen, Schlatter, Ivanell, Sørensen, and Henningson]{sarmast_mutual_2014}
S.~Sarmast, R.~Dadfar, R.~F. Mikkelsen, P.~Schlatter, S.~Ivanell, J.~N. Sørensen, and D.~S. Henningson.
\newblock Mutual inductance instability of the tip vortices behind a wind turbine.
\newblock \emph{Journal of Fluid Mechanics}, 755:\penalty0 705--731, Sept. 2014.
\newblock ISSN 0022-1120, 1469-7645.
\newblock \doi{10.1017/jfm.2014.326}.
\newblock URL \url{https://www.cambridge.org/core/product/identifier/S0022112014003267/type/journal_article}.

\bibitem[Schmid(2010)]{schmid_dynamic_2010}
P.~J. Schmid.
\newblock Dynamic mode decomposition of numerical and experimental data.
\newblock \emph{Journal of Fluid Mechanics}, 656:\penalty0 5--28, Aug. 2010.
\newblock ISSN 1469-7645, 0022-1120.
\newblock \doi{10.1017/S0022112010001217}.
\newblock URL \url{https://www.cambridge.org/core/journals/journal-of-fluid-mechanics/article/dynamic-mode-decomposition-of-numerical-and-experimental-data/AA4C763B525515AD4521A6CC5E10DBD4}.
\newblock Publisher: Cambridge University Press.

\bibitem[Schnipper et~al.(2009)Schnipper, Andersen, and Bohr]{schnipper_vortex_2009}
T.~Schnipper, A.~Andersen, and T.~Bohr.
\newblock Vortex wakes of a flapping foil.
\newblock \emph{Journal of Fluid Mechanics}, 633:\penalty0 411--423, Aug. 2009.
\newblock ISSN 1469-7645, 0022-1120.
\newblock \doi{10.1017/S0022112009007964}.
\newblock URL \url{https://www.cambridge.org/core/journals/journal-of-fluid-mechanics/article/vortex-wakes-of-a-flapping-foil/66F46A94D1181911512F092735C2BDCC}.
\newblock Publisher: Cambridge University Press.

\bibitem[Smits(2019)]{smits_undulatory_2019}
A.~J. Smits.
\newblock Undulatory and oscillatory swimming.
\newblock \emph{Journal of Fluid Mechanics}, 874:\penalty0 P1, Sept. 2019.
\newblock ISSN 0022-1120, 1469-7645.
\newblock \doi{10.1017/jfm.2019.284}.
\newblock URL \url{https://www.cambridge.org/core/journals/journal-of-fluid-mechanics/article/undulatory-and-oscillatory-swimming/74B7173131202AE623DFD70215487D30}.
\newblock Publisher: Cambridge University Press.

\bibitem[Strom et~al.(2022)Strom, Polagye, and Brunton]{strom_near-wake_2022}
B.~Strom, B.~Polagye, and S.~L. Brunton.
\newblock Near-wake dynamics of a vertical-axis turbine.
\newblock \emph{Journal of Fluid Mechanics}, 935:\penalty0 A6, Mar. 2022.
\newblock ISSN 0022-1120, 1469-7645.
\newblock \doi{10.1017/jfm.2021.1123}.
\newblock URL \url{https://www.cambridge.org/core/journals/journal-of-fluid-mechanics/article/nearwake-dynamics-of-a-verticalaxis-turbine/A2F6F94238D6CA901ED33B863C257266}.
\newblock Publisher: Cambridge University Press.

\bibitem[Taira et~al.(2020)Taira, Hemati, Brunton, Sun, Duraisamy, Bagheri, Dawson, and Yeh]{taira_modal_2020}
K.~Taira, M.~S. Hemati, S.~L. Brunton, Y.~Sun, K.~Duraisamy, S.~Bagheri, S.~T.~M. Dawson, and C.-A. Yeh.
\newblock Modal {Analysis} of {Fluid} {Flows}: {Applications} and {Outlook}.
\newblock \emph{AIAA Journal}, 58\penalty0 (3):\penalty0 998--1022, Mar. 2020.
\newblock ISSN 0001-1452, 1533-385X.
\newblock \doi{10.2514/1.J058462}.
\newblock URL \url{https://arc.aiaa.org/doi/10.2514/1.J058462}.

\bibitem[Taylor(2018)]{taylor_simple_2018}
G.~K. Taylor.
\newblock Simple scaling law predicts peak efficiency in oscillatory propulsion.
\newblock \emph{Proceedings of the National Academy of Sciences}, 115\penalty0 (32):\penalty0 8063--8065, Aug. 2018.
\newblock ISSN 0027-8424, 1091-6490.
\newblock \doi{10.1073/pnas.1809769115}.
\newblock URL \url{https://pnas.org/doi/full/10.1073/pnas.1809769115}.

\bibitem[Taylor(1997)]{taylor1997introduction}
J.~Taylor.
\newblock \emph{An introduction to error analysis: the study of uncertainties in physical measurements}.
\newblock University Science Books, 1997.

\bibitem[Triantafyllou et~al.(1993)Triantafyllou, Triantafyllou, and Grosenbaugh]{triantafyllou_optimal_1993}
G.~S. Triantafyllou, M.~S. Triantafyllou, and M.~A. Grosenbaugh.
\newblock Optimal {Thrust} {Development} in {Oscillating} {Foils} with {Application} to {Fish} {Propulsion}.
\newblock \emph{Journal of Fluids and Structures}, 7\penalty0 (2):\penalty0 205--224, 1993.
\newblock ISSN 0889-9746.
\newblock \doi{https://doi.org/10.1006/jfls.1993.1012}.
\newblock URL \url{https://www.sciencedirect.com/science/article/pii/S0889974683710121}.

\bibitem[Tu et~al.(2014)Tu, Rowley, Luchtenburg, Brunton, and Kutz]{tu_dynamic_2014}
J.~H. Tu, C.~W. Rowley, D.~M. Luchtenburg, S.~L. Brunton, and J.~N. Kutz.
\newblock On dynamic mode decomposition: {Theory} and applications.
\newblock \emph{Journal of Computational Dynamics}, 1\penalty0 (2):\penalty0 391--421, Nov. 2014.
\newblock ISSN 2158-2491.
\newblock \doi{10.3934/jcd.2014.1.391}.
\newblock URL \url{https://www.aimsciences.org/en/article/doi/10.3934/jcd.2014.1.391}.
\newblock Publisher: Journal of Computational Dynamics.

\bibitem[Tytell and Lauder(2004)]{tytell_hydrodynamics_2004}
E.~D. Tytell and G.~V. Lauder.
\newblock The hydrodynamics of eel swimming.
\newblock \emph{Journal of Experimental Biology}, 207\penalty0 (11):\penalty0 1825--1841, May 2004.
\newblock ISSN 1477-9145, 0022-0949.
\newblock \doi{10.1242/jeb.00968}.
\newblock URL \url{https://journals.biologists.com/jeb/article/207/11/1825/14825/The-hydrodynamics-of-eel-swimmingI-Wake-structure}.

\bibitem[Zhang(2017)]{zhang_footprints_2017}
J.~Zhang.
\newblock Footprints of a flapping wing.
\newblock \emph{Journal of Fluid Mechanics}, 818:\penalty0 1--4, May 2017.
\newblock ISSN 0022-1120, 1469-7645.
\newblock \doi{10.1017/jfm.2017.173}.
\newblock URL \url{https://www.cambridge.org/core/journals/journal-of-fluid-mechanics/article/footprints-of-a-flapping-wing/922F74C74132247F93DE2E4FF9529481}.
\newblock Publisher: Cambridge University Press.

\end{thebibliography}
\end{document}